\documentclass[11pt,a4paper]{article}
\usepackage{jheppub} 
\usepackage{bm}
\usepackage{slashed,bbold}

\newcommand{\beq}{\begin{equation}} 
\newcommand{\eeq}{\end{equation}}
\newcommand{\beqa}{\begin{eqnarray}} 
\newcommand{\eeqa}{\end{eqnarray}}

\newcommand{\Tr}{{\textrm {Tr}}}

\renewcommand{\d}[1]{\ensuremath{\operatorname{d}\!{#1}}}

\title{Singular perturbation theory for the  thermodynamic properties of holographic QCD}

\author{Eugenio Meg\'{i}as,}
\author{Manuel Valle}

\affiliation{Departamento de F\'{\i}sica Te\'orica, 
Universidad del Pa\'{\i}s Vasco UPV/EHU, \\
Apartado 644, 48080 Bilbao, Spain}

\emailAdd{eugenio.megias@ehu.eus}
\emailAdd{manuel.valle@ehu.es}

\abstract{We explore the thermodynamics of a black-hole solution
in  improved holographic QCD with a simple dilaton potential having two parameters. By applying techniques of singular perturbation theory, we get uniform approximations for the metric and the dilaton field in the two regimes of big and small black-holes. These techniques lead to a resummation of the naive expansion at high temperatures, providing an important theoretical improvement with respect to previous results in the literature. By using this technique, it is shown how a quadratic dependence at low enough temperatures can naturally appear in the free energy. A comparison with lattice data of gluodynamics is performed. It is provided as well an estimate of the value of the gluon condensate at zero temperature which turns out to be in quite good agreement with the accepted values in the literature from phenomenological studies of QCD.
}

\begin{document}
\maketitle
\flushbottom

%%%%%%%%%%%%%%%%%%%%%%%%%%%%%%%%%%%%%%%%%%%%%%%%%%%%%%%%%%%%%%%%%%%%%%%%%%%%%%%%%%%%%%%%%%%%%%%%%%%%%%%%%%%%%%%%%%%%%%%%%%%%%%%%%%%%%%

\section{Introduction}
\label{sec:introduction}

The gauge/gravity duality is nowadays a powerful tool to study the properties of gauge theories, and in particular of QCD, in their strongly coupled regime either at zero or finite temperature. One of the most important applications of this duality 
is the physics of strongly coupled plasmas. In particular, we can study the thermodynamics of a field theory from the classical computation of the thermodynamics of black holes in the gravity dual. This duality can be expressed in the form  
\begin{equation}
  S_{\textrm{Black Hole}}(T) = \frac{{\cal A}(r_{\textrm{horizon}})}{4 G_D} \longleftrightarrow S_{\textrm{QCD}}(T) \,.
\end{equation}
The entropy of a black hole can be obtained classically from the 
famous Bekenstein-Hawking entropy formula, where ${\cal A}(r_{\textrm{horizon}})$
 is the area of the black hole horizon, see e.g.~\cite{Gubser:2008ny, Gursoy:2008za,Alanen:2009xs,Megias:2010ku,Li:2011hp}. 
In conformal AdS$_5$ the metric has a horizon in the 
bulk space at $r_h = \pi \ell^2 T$ where $\ell$ is the radius of the AdS space, 
and the entropy scales like $S_{\textrm{Black Hole}} \propto r_h^3 \propto T^3$. 
However, in order to have a reliable extension of this duality 
to SU($\text{N}_c$) Yang-Mills theory, the first task is to control the breaking of conformal invariance. 

Gluodynamics is invariant under scale and conformal transformations at the classical level, but this classical invariance is broken by quantum corrections due to the necessary regularization of the UV divergences. This yields the so-called {\it trace anomaly}~\cite{Collins:1976yq}, corresponding to the divergence of the dilatation current which is equal to the trace of the energy-momentum tensor $T^\mu_\mu$~\cite{Callan:1970ze}.  At finite temperature, the energy density~$\varepsilon$ and the pressure $p$ enter as~\cite{Landsman:1986uw}
\begin{equation}
\langle T^\mu_\mu\rangle = -\varepsilon + 3p = -\frac{\beta(g)}{2g^3} \langle (F_{\mu\nu}^a)^2\rangle 
\end{equation}
in the mostly plus convention, where $F_{\mu\nu}^a$ is the field strength tensor and $\beta = \mu \partial g / \partial \mu$ is the beta function. Only in an ideal gas one has $\varepsilon = 3p$ and $\langle T^\mu_\mu \rangle = 0$, so that a non-vanishing value of the trace anomaly provides a measure of the departure from conformality and, equivalently, of the strength of the interaction between the constituents of the systems.

The equation of state of QCD has been studied for a long time by using different methods. A naive weak coupling expansion turns out to be poorly convergent in the regime of temperatures close to the phase transition, see e.g.~\cite{Braaten:1995jr,Kajantie:2002wa}. It has been proposed in the literature several methods to resum the perturbative expansion, one of the most popular being the Hard Thermal Loop (HTL), currently computed up to 3-loops order~\cite{Andersen:2009tc}. However, all these methods fail to reproduce the lattice data for the trace anomaly in the regime $ T_c \lesssim T \lesssim (2.5  - 3) T_c$, which corresponds to a strongly interacting quark-gluon plasma picture.

There are many works focusing on the computation of the equation of state of QCD in lattice, see e.g.~\cite{Boyd:1996bx,Borsanyi:2012ve}. As it has been shown in~\cite{Pisarski:2006yk,Megias:2009mp,Panero:2009tv,Megias:2009ar}, these lattice data show a clear behavior close to $T_c$ of the form
\begin{equation}
\frac{\varepsilon-3p}{T^4} = a(\mu_T) + b \left(\frac{T_c}{T}\right)^2 \,, \label{eq:e3pfit}
\end{equation}
leading to the existence of power corrections in $T^{-2}$. This behavior contradicts perturbation theory which contains no powers but only logarithms in the temperature, a feature shared by HTL and other resummation techniques. This formula can be understood as a separation of perturbative contributions $a(\mu_T) \sim 1/\log T$ which are generically small, and non-perturbative ones characterized by power corrections. There are other observables in QCD with similar patterns including power corrections contributions, like for instance the Polyakov loop~\cite{Megias:2005ve,Andreev:2009zk} or the heavy quark-antiquark free energy~\cite{Megias:2007pq}. It was conjectured in~\cite{Narison:2009ag} the existence of a duality between large order QCD perturbative series and non-perturbative power corrections, stating that the latter can be obtained from the former after considering high enough order series. This duality was studied in~\cite{Megias:2009ar} at finite temperature from an analysis of the statistical correlations between the two terms in the rhs of Eq.~(\ref{eq:e3pfit}). In the present work we will study analytically the equation of state of an improved holographic model for QCD in an expansion at high enough temperatures. As we will see, this technique leads to an implicit resummation of the perturbative series, and as a side effect a remnant of the power corrections in $T^2$ already appears in the result. Some numerical results will be presented as well.

The manuscript is organized as follows. We introduce in Sec.~\ref{sec:model} the holographic model, derive the relevant equations of motion and provide the exact zero temperature solution. In Sec.~\ref{sec:Resummation} we explain the technique to perform a resummation of the solution of the equations of motion in the two opposite regimes $\Phi_h \to \pm\infty$, where $\Phi_h$ is the scalar field at the horizon. In Sec.~\ref{sec:thermodynamics} it is obtained the equation of state, and discussed the thermodynamic consistency of the result. The equation of state is then used in Sec.~\ref{sec:lattice} to perform a comparison with the lattice data of gluodynamics. Finally, we discuss in Sec.~\ref{sec:TwoVacua} some physical consequences of a kind of RG flows connection two fixed points, and provide an estimate for the gluon condensate at zero temperature. We conclude with a discussion of the results, and an outlook towards future directions in Sec.~\ref{sec:discussion}.

%---------------------------------------------------------------------------------
%---------------------------------------------------------------------------------
\section{The Improved Holographic QCD model}
\label{sec:model}

The bottom-up approach turns out to be quite useful to study the thermodynamics of QCD in the strongly coupled regime. It is based on the building of a gravity dual of QCD, including the main properties of this theory. We introduce in this section the model, and provide the relevant equations of motion.

\subsection{The model}
\label{subsec:model}

One of the most successful models within the bottom-up scenario is the 
5D Einstein-dilaton model, with the Euclidean action~\cite{Gursoy:2008za}
\begin{equation}
\begin{split}
\label{eq:Smodel}
S &= \frac{1}{2\kappa^2} \int d\rho \, d^4x \sqrt{G} \Bigl(-R + G^{\mu\nu} \partial_\mu \Phi \partial_\nu \Phi + 2V(\Phi) \Bigr) \nonumber \\
&\quad -\frac{1}{\kappa^2} \int_{\rho = \epsilon} d^4x \sqrt{h}\, K  \,,
\end{split} 
\end{equation}
where $\kappa^2$ is the 5D Newton constant, $\Phi$~is a scalar field
to be identified with the Yang-Mills coupling through $g^2 = e^{\gamma
  \Phi}$, and $\epsilon$ is a cut-off surface near the boundary.
Taking the limit $\epsilon \to 0$, one takes the surface to the
AdS$_5$ boundary.  The boundary term is the usual Gibbons-Hawking
contribution built up from the extrinsic curvature $K$ and the
determinant~$h$ of the induced metric at the boundary. The
introduction of a scalar field breaks conformal invariance, and the
form of the scalar potential $V(\Phi)$ is usually phenomenologically
adjusted to describe some observables of QCD, like for instance the
trace anomaly.  There are in the literature many different proposals
for the dilaton potential.  In this work we will consider the form
\begin{equation}
\begin{split}
\label{eq:Vphi}
 V(\Phi) &= -\frac{6}{\ell^2} \left( 1 + \frac{2v_0}{3} e^{\gamma\Phi} + 
    \frac{v_0^2}{36}(4-3\gamma^2) e^{2\gamma\Phi} \right) \\ 
    &= \frac{1}{2} W'(\Phi)^2 - \frac{2}{3} W(\Phi)^2 \,, 
\end{split}
\end{equation}
generated by the simple superpotential $W(\Phi) = \ell^{-1}(3 + v_0
e^{\gamma \Phi})$.  $\ell$ is the radius of the asymptotically AdS$_5$
background.  The combination $\gamma^2 v_0/2 \equiv b_0$ will be taken
from the one-loop $\beta$-function
\begin{equation}
\label{eq:beta}
\mu \frac{dg}{d\mu} = -b_0 g^3 =  -\frac{11 \, \text{N}_c}{3 (4 \pi)^2} g^3
\end{equation}
of a pure SU($\text{N}_c$) gauge theory. 

We will study finite temperature solutions of the Einstein-scalar model corresponding to a black hole of the 
form
 \begin{equation}
 \label{eq:metric}
 ds^2 = G_{\mu\nu} dx^\mu dx^\nu = \frac{\ell^2}{4\rho^2} d\rho^2 + \frac{\ell^2}{\rho} g_{\tau\tau}(\rho) d\tau^2 +  \frac{\ell^2}{\rho} g_{xx}(\rho) d{\vec{x}}^2 \,, 
 \end{equation}
with a regular horizon at $\rho=\rho_h$, i.e. $g_{\tau\tau}(\rho_h)= g_{\tau\tau}'(\rho_h)=0$. 
The horizon data define the temperature and entropy density through  
\begin{equation}
\label{eq:TS}
\begin{split}
T &= \frac{1}{2\pi} \sqrt{2\rho_h g_{\tau\tau}^{\prime\prime}(\rho_h)} \, ,  \\
s &= \frac{2\pi}{\kappa^2} \left( \frac{\ell^2}{\rho_h } g_{xx}(\rho_h) \right)^{3/2} .
\end{split}
\end{equation}
We work in Fefferman-Graham coordinates as these allow for a clean decoupling of the scalar field equation.  
Also, the asymptotic expansions for the metric and the dilaton field in models of  Improved Holographic QCD
have been derived in these coordinates~\cite{Papadimitriou:2011qb}.

\subsection{Equations of motion}
\label{subsec:eom}

A nice property of the field equations following from Eq.~(\ref{eq:Smodel}) 
is that the equation for the scalar field is decoupled by 
an additional differentiation. 
The third order equation that results is  equi-dimensional, 
\begin{eqnarray}
\label{eq:phi}
\begin{split}
0 &= 48 \rho^4 \Phi' \Phi''' - 96 \rho^4 \Phi''^2   + \bigl(36 \ell^2  V'(\Phi) - 48 \rho \Phi' \bigr) \rho^2 \Phi''   \\ 
 & \quad + \bigl(-48 - 32 \ell^2 V(\Phi) - 12 \ell^2 V''(\Phi) \bigr) \rho^2 \Phi'^2 \\ 
 & \quad + 36 \ell^2 V'(\Phi)  \rho \Phi' -3 \ell^4 \bigl(V'(\Phi) \bigr)^2  \,, 
\end{split}
\end{eqnarray}
and  is converted in autonomous by making the  change of variables, 
$(\rho, \Phi) \to (\Phi, u(\Phi))$, where  $u(\Phi) \equiv \rho \Phi'(\rho)$: 
\begin{equation}
\label{eq:u}
\begin{split}
u''(\Phi) &= \frac{\bigl(u'(\Phi) \bigr)^2}{u(\Phi)} - \frac{3 \ell^2 V'(\Phi)}{4 u(\Phi)^2} u'(\Phi) + 
\frac{\ell^4 \bigl(V'(\Phi) \bigr)^2}{16 u(\Phi)^3} + \frac{\ell^2 \bigl(8 V(\Phi) + 3 V''(\Phi) \bigr)}{12 u(\Phi)} \,.
\end{split}
\end{equation}
The remaining  field equations written in terms of the scalar field are
\begin{align}
u(\Phi)^2 g_{xx}''(\Phi) + u(\Phi)(u'(\Phi)-2) g_{xx}'(\Phi) + \left(1 + \frac{\ell^2}{6} V(\Phi) + \frac{1}{3}  u(\Phi)^2 \right) g_{xx}(\Phi) &= 0 \,,  \label{eq:gxxu} \\ 
g_{\tau \tau}'(\Phi)  +
  g_{\tau \tau}(\Phi) \bigg( \frac{ g_{xx}'(\Phi)}{g_{xx}(\Phi)}  -\frac{2}{u(\Phi)} + \frac{g_{xx}(\Phi) (-\ell^2 V(\phi) + 
 2 u(\Phi)^2)}{3 u(\Phi) \bigl(g_{xx}(\Phi)  - u(\Phi) g_{xx}'(\Phi) \bigr) }\bigg) &= 0 \,. \label{eq:gttu}
\end{align}

\subsection{Zero temperature solution}
\label{subsec:ZeroTemperature}

At zero temperature, when $g_{xx} = g_{\tau \tau}$, 
the solution corresponds  to a domain wall configuration, 
$u_0(\Phi) = \frac{\ell}{2} W'(\Phi) = \tfrac{\gamma v_0}{2} e^{\gamma \Phi}$. It  is given by 
\begin{equation}
\label{eq:zeroim}
\begin{split}
e^{\gamma \Phi(\rho)} &= \frac{2}{ \gamma^2 v_0\,  \log \left(\rho_0/\rho \right) } \,,  \\ 
g_{xx}(\Phi) &= g_{\tau \tau}(\Phi) = g_{(0)} e^{-2 \Phi/(3 \gamma)}  \,, 
\end{split}
\end{equation}
where $\rho_0$ is an integration constant defining the location of a
singularity at which $\Phi = \infty$.  If we identify $\gamma^2 v_0/2$
with $b_0$ and $g^2 = e^{\gamma \Phi}$, it turns out that the radial
coordinate $\rho^{-1/2}$ may be clearly interpreted as the RG scale,
and $\rho_0^{-1/2} \equiv \Lambda_{QCD}$ corresponds to the location
of the Landau pole of QCD.  The other integration constant $g_{(0)}$
is arbitrary, and it will determine the leading asymptotic behavior of the
metric as $\Phi \to -\infty$.  Notice the exponential grow of the
metric near the boundary, unlike the behavior $g_{ij} \sim 1$ for the
superpotential $W(\Phi) =3 \ell^{-1}$.

This solution with Poincar\'e invariance can be written equivalently as a domain wall, 
\begin{equation}
\label{eq:zeroDW}
\begin{split}
\d s^2 &= dr^2 + e^{2 A(r)} (d\tau^2 + d{\vec x}^2) ,  \quad A'(r) = \frac{1}{3} W(\Phi) , 
\quad \Phi'(r) = -W'(\Phi)  \,, \\   
e^{\gamma \Phi} &= \frac{\ell}{\gamma^2 v_0 r}, \\ 
e^{2 A} &= e^{2 r/L} \left(\frac{r}{\ell} \right)^{2/(3 \gamma^2)}  \bar{g}_{(0)} \,. 
\end{split}
\end{equation} 
Given that  the system is autonomous, the other constant of integration has been chosen to 
have $\Phi=\infty$ at $r=0$. This solution can be connected with the solution of Eq.~(\ref{eq:zeroim}) by assuming that the relation between $\rho$ and $r$ is 
\begin{equation}
\label{eq:conv}
r = \frac{\ell}{2} \log \frac{\rho_0}{\rho} = -\frac{\ell}{2} \log \rho \Lambda_{QCD}^2  ,  \quad 0 < \rho \leq \Lambda_{QCD}^{-2} \,. 
\end{equation}

\section{Asymptotically AdS black hole solutions from resummation}
\label{sec:Resummation}

We now consider a black hole solution specified by the horizon data $(\rho_h, \Phi_h)$. Local analysis of Eqs.~(\ref{eq:u}), (\ref{eq:gxxu})
and~(\ref{eq:gttu}) shows that, in order to have a regular horizon at
$\Phi= \Phi_h$, it is necessary that $\lim_{\Phi \to \Phi_h}
\bigl(u(\Phi) g'_{xx}(\Phi) - g_{xx}(\Phi)\bigr) = 0$.  This condition
serves to fix the first term of power series solutions with movable singularities
\begin{align}
\label{eq:seriesu}
u(\Phi) &= \sum_{n=0}^\infty a_n(\Phi_h) (\Phi_h - \Phi)^{n + 1/2} \,,   \\ 
g_{xx}(\Phi) &= g_{xx}(\Phi_h)\Bigl(1 + \sum_{n=1}^\infty b_n(\Phi_h) (\Phi_h - \Phi)^{n/2} \Bigr) \,, \label{eq:seriesgxx} \\ 
g_{\tau \tau}(\Phi) &= -g_{\tau \tau}'(\Phi_h)\Bigl(\Phi_h - \Phi + \sum_{n=3}^\infty d_n(\Phi_h) (\Phi_h - \Phi)^{n/2} \Bigr)  \,, \label{eq:seriesgtt}
\end{align}
when one makes the replacement of these ansatzs in Eqs.~(\ref{eq:u})-(\ref{eq:gttu}). The results for the first terms are
\begin{equation}
\label{eq:a0_b1}
\begin{split}
a_0 &= \frac{\sqrt{e^{\gamma \Phi_h} v_0 \gamma \bigl(12 + 
   e^{\gamma \Phi_h} v_0 (4 - 3 \gamma^2) \bigr)    } }{2 \sqrt{3} } \,,  \\ 
a_1 &= \frac{-144 + v_0 e^{\gamma \Phi_h}\bigl(v_0 e^{\gamma \Phi_h} (-4 + 3 \gamma^2)(4+ 9 \gamma^2) -
  6(16 + 9 \gamma^2) \bigr) }{24 \sqrt{3} \sqrt{e^{\gamma \Phi_h} v_0 \gamma \bigl(12 + 
   e^{\gamma \Phi_h} v_0 (4 - 3 \gamma^2) \bigr)}} \,, \\ 
b_1 &= d_3= -\frac{4 \sqrt{3}}{\sqrt{e^{\gamma \Phi_h} v_0 \gamma \bigl(12 + 
   e^{\gamma \Phi_h} v_0 (4 - 3 \gamma^2) \bigr)    } }  \,, 
\end{split}
\end{equation} 
and the remainder are determined recursively. The horizon data  $g_{xx}(\Phi_h)$ and  $g_{\tau \tau}'(\Phi_h)$ are still to be determined. 

The special case in which the square root of Eqs.~(\ref{eq:a0_b1}) vanishes occurs  
when $V'(\Phi_{min}) =0$, where we have replaced $\Phi_h$ by $\Phi_{min}$ in these equations. Here $\Phi_{min}$  is the value of a critical point of the potential that appears only for $\gamma > \gamma_c \equiv 2/\sqrt{3}$. It is understood that now $\Phi_{min}$ does not have the meaning of a horizon. Indeed, the position of a possible horizon must be below $\Phi_{min}$. This  solution corresponds to a domain wall interpolating between $\Phi=-\infty$ and $\Phi_{min}$ (see also Section~\ref{sec:TwoVacua}). Now we have $u_c(\Phi) = \tfrac{\ell}{2} W_c' (\Phi)$, where $W_c(\Phi)$ is the solution of Eq.~(\ref{eq:Vphi}),  which near the critical points behaves as
\begin{equation}
\label{eq:Wc}
 W_c(\Phi) \sim \begin{cases}
 \frac{1}{\ell} (3  + v_0 e^{\gamma \Phi} ) , & \Phi \to -\infty  \\ 
 \frac{3 \gamma}{\ell \sqrt{\gamma^2 - 4/3}} - 
 \frac{\gamma(\sqrt{5}-1)}{\ell \sqrt{\gamma^2 - 4/3} } (\Phi_{min} - \Phi)^2 \,, & \Phi \to \Phi_{min} \,.
  \end{cases}
\end{equation}

In the rest of this section we will study the analytical solutions in the regimes: i)~$\Phi_h \gg 1$ with $\gamma < 2/\sqrt{3}$, and ii) $\nu \equiv \gamma v_0 e^{\gamma \Phi_h} \ll 1$  for any $\gamma >0$.

\subsection{Resummation when $\Phi_h \gg 1$}
\label{subsec:Phihlarge}

By simple inspection one sees that when $\Phi_h \gg 1$, the leading part of all the coefficients $a_n$ in Eq.~(\ref{eq:a0_b1}) is proportional to $e^{\gamma \Phi_h}$. It is relatively easy to make the resummation of the series with their coefficients  approximated in this way.  By making the ansatz $u(\Phi) = e^{\gamma \Phi_h} R(\Phi_h - \Phi)$ in Eq.~(\ref{eq:u}), and keeping the leading terms when $\Phi_h \to  \infty$, one obtains a simpler equation that may be integrated to give the desired solution. 
This is given by
\begin{equation}
\label{eq:unif3}
 u(\Phi) = \frac{v_0 \gamma}{2} e^{\gamma \Phi}   \sqrt{1 - e^{-q (\Phi_h - \Phi)} } + \ldots \,,
\end{equation} 
with $q \equiv (4 - 3 \gamma^2)/(3 \gamma)$, which requires $q >0$. Note that this expression is a good approximation  for any value of the scalar field when $\Phi_h \to \infty$ since, for large $\Phi_h - \Phi$, it goes to the zero temperature solution. The same procedure may be applied to Eqs.~(\ref{eq:gxxu}) and~(\ref{eq:gttu}) to determine solutions which uniformly approximate  the metric over  all range:
\begin{equation}
\begin{split}
g_{xx}(\Phi) &= g_{xx}(\Phi_h) e^{-2(\Phi - \Phi_h)/(3 \gamma)} \left( 1 - 
\frac{12 e^{-\gamma \Phi_h} \sqrt{1 - e^{-q (\Phi_h - \Phi)} }}{v_0 (4-3 \gamma^2)} \right) ,  \\ 
g_{\tau \tau}(\Phi)&= g_{xx}(\Phi_h) e^{-2(\Phi - \Phi_h)/(3 \gamma)}\left( 
1-  e^{-q (\Phi_h - \Phi)} \right) \,. 
\end{split}
\end{equation}
The matching with the zero temperature solution of Eq.~(\ref{eq:zeroim}) then produces the relations
\begin{align}
\label{eq:g1der}
g_{\tau \tau}'(\Phi_h) &=  -g_{(0)} q \, e^{-2 \Phi_h/(3  \gamma)}  \,,  \\ 
\label{eq:gxder}
g_{xx}(\Phi_h) &= g_{(0)} e^{-2 \Phi_h/(3  \gamma)} ,  \qquad \Phi_h \gg 1 \,. 
\end{align}

\subsection{Boundary-layer analysis when $\nu \equiv \gamma v_0 e^{\gamma \Phi_h} \ll 1$}
\label{subsec:BoundaryLayer}

An extreme opposite regime can be addressed as follows.  For
negatively large values of the dilaton at the horizon, or small values
of $v_0$, the leading part of each coefficient of the series
(\ref{eq:seriesu}) behaves as $a_n(\Phi_h) \propto \nu^{-n +
  1/2}$. This suggests to make the replacement $u^{\mathrm{in}}(\Phi)
= \nu\, U_1\bigl((\Phi_h -\Phi)/\nu \bigr)$ in Eq.~(\ref{eq:u}) in order
to account for this behavior and include great variations near the
horizon. In fact, this is the first non-zero term of an inner
expansion $u^{\mathrm{in}}(\Phi) = \sum_{n=1}^\infty \nu^n U_n((\Phi_h
- \Phi)/\nu)$, valid in the region of boundary-layer near the horizon,
where the inner variable, $\Psi \equiv (\Phi_h - \Phi)/\nu$, is
$O(1)$. We have determined the two first orders of this expansion
which reads
\begin{equation}
\label{eq:inner1}
\begin{split}  
u^{\mathrm{in}}(\Phi) &= \frac{\nu}{2}  \sqrt{1 - e^{- 4 \Psi}} \\ 
&\quad + \frac{\nu^2 \, \gamma e^{-4 \Psi}}{2 \sqrt{1 - e^{- 4 \Psi}}} \Bigl[\frac{\pi^2}{48} - 
 \frac{1}{8} \text{Li}_2 \bigl(e^{-4 \Psi} \bigr) + \Psi^2  \\ 
& \quad + \Bigl(\frac{2}{3 \gamma^2} - e^{4 \Psi} +  \log \sqrt{1 - e^{-4 \Psi} } \Bigr) \Psi  \Bigr] 
+ O(\nu^3) \,, 
\end{split}
\end{equation} 
where $\text{Li}_2$ is the dilogarithmic function. For the regime outside the boundary layer we tried a solution of the form $u^\mathrm{out}(\Phi) = \sum_{n=1}^\infty   \nu^n u_n(\Phi_h - \Phi)$. The corresponding substitution in Eq.~(\ref{eq:u})  leads to $u^\mathrm{out}(\Phi)  = \frac{\nu}{2} e^{\gamma (\Phi - \Phi_h)}$ to all orders in $\nu$, so that the outer solution has the same form as the zero-temperature solution. Thus the asymptotic matching produces a uniform approximation valid for $(-\infty, \Phi_h)$  given by   
\begin{equation}
\label{eq:unif}
\begin{split}
u^{\mathrm{unif}}(\Phi) &= \frac{v_0 \gamma}{2} e^{\gamma \Phi} + u^{\mathrm{in}}(\Phi) -\frac{v_0 \gamma}{2} e^{\gamma \Phi_h} + \frac{\nu^2 \gamma}{2}\frac{\Phi_h - \Phi}{\nu} + O(\nu^3) \,, 
\end{split}
\end{equation} 
where the two last terms are the common limit of the inner and outer approximation in the matching region, $\nu \ll \Phi_h - \Phi \ll 1$.  
Now we have $|u(\Phi) - u^{\mathrm{unif}}(\Phi)| =  O(\nu^3)$ in $(-\infty, \Phi_h)$. 

This analysis does not include the complete asymptotics of $u(\Phi)$  
in the UV (outer) region, but this can be derived from the matching with the inner expansion. The solution of  Eq.~(\ref{eq:u}) near the boundary  shows that the subleading behavior of $u$ when $\Phi \to -\infty$  has a term, not included in the outer expansion,  
with a non-analytical dependence on $\nu$ of the form
\begin{equation}
\delta u(\Phi) = C_\Phi  \exp\left[-4 e^{-\gamma \Phi}/(\gamma^2  v_0) + (4/(3 \gamma) - \gamma)\Phi \right]
\left( 1 + \frac{v_0 \, e^{\gamma \Phi}}{3} \right) \,,  \label{eq:deltau}
\end{equation} 
where $C_\Phi$ is a quantity depending only on $\Phi_h$.   
The matching with the inner solution of Eq.~(\ref{eq:inner1})  yields the $C_\Phi$-coefficient: 
\begin{equation}
\label{eq:Cphi}
\begin{split}
C_\Phi(\Phi_h) &=  \exp \left(\frac{4}{\gamma  \nu} - 4 \Phi_h/(3  \gamma) + \gamma \Phi_h  \right) \times \left( -\frac{\nu}{4} + \frac{\nu^2 (8 + \gamma^2 \pi^2) }{96 \gamma}  + O(\nu^3) \right) .
\end{split}
\end{equation}
This coefficient is very important because it is related to the QCD trace anomaly, as we will see below.

Now from the knowledge of the two first orders of $u^{\mathrm{in}}(\Phi)$, 
it is possible to get in closed form the two lowest orders in $\nu$ for the inner solutions of the metric, 
which have the form  $g_{i j}^{\mathrm{in}}(\Phi) = \sum_n \nu^n g_{(n) i j}(\Psi)$. For simplicity, we only write the leading term of these expansions. They read 
\begin{equation}
\begin{split}
g^{\mathrm{in}}_{xx}(\Phi) &= g_{xx}(\Phi_h) e^{4 \Psi} \left(1 -  \sqrt{1 - e^{- 4 \Psi} } \right) \,,  \\ 
g^{\mathrm{in}}_{\tau \tau}(\Phi) &= -\frac{g_{\tau \tau}'(\Phi_h)\,  \nu  \sinh\bigl(2\Psi \bigr)}{2 \sqrt{
-1+2 e^{4 \Psi} \bigr(1 + \sqrt{1 - e^{- 4 \Psi} } \bigl)} } \,.
\end{split}
\end{equation}
Again, the part of the outer solution of  the form $\sum_{n=0}^\infty \nu^n g_{i j \,(n)} (\Phi_h- \Phi)$ 
reduces to the zero-temperature result  $g_{(0)} e^{-2 \Phi/(3  \gamma)}$.  
The asymptotic matching of this outer solution with the two first orders in $\nu$ 
of the inner expansion determines the horizon quantities 
$g_{xx}(\Phi_h)$ and $g_{\tau \tau}'(\Phi_h)$,  
which are related to the temperature and entropy of the black hole.~\footnote{
With abuse of notation, the second derivative in Eq.~(\ref{eq:TS}) is written as 
$g_{\tau \tau}''(\rho_h) =    
\rho_h^{-2} \lim_{\Phi \to \Phi_h} \bigl(u(\Phi) u'(\Phi) g_{\tau \tau}'(\Phi) \bigr)$.} 
They take the form
\begin{align}
\label{eq:der1}
g_{\tau \tau}'(\Phi_h) &= 
 -\frac{8 g_{(0)} e^{-2 \Phi_h/(3 \gamma)}}{\nu}  \left( 1 + \frac{\nu}{12 \gamma} (4 + 3\gamma^2 - 4\log 2)   + O(\nu^2)  \right) , \\ 
\label{eq:ders}
g_{x x}(\Phi_h) &=  2 g_{(0)} e^{-2 \Phi_h/(3  \gamma)} \left(1 -\frac{\nu \, \log 2}{3 \gamma} +  O(\nu^2)  \right) . 
\end{align}

Finally, local analysis of Eqs.~(\ref{eq:gxxu}) and~(\ref{eq:gttu}) when $\Phi \to -\infty$ shows that the sub-leading  parts of the outer metric have an expansion in  series of negative powers of $\nu$ that read 
\begin{align}
\delta g_{xx}(\Phi) & = \exp\left[-4 e^{-\gamma \Phi}/(\gamma^2  v_0) + (2/(3 \gamma) - \gamma)\Phi \right] \nonumber \\  
& \quad \times 
 \biggl( \frac{g_{(0)}}{3 \gamma} C_\Phi 
  + \frac{v_0 g_{(0)} e^{\gamma \Phi} }{36 \gamma} \bigl( C_\Phi (3 \gamma^2 - 4) - 9 \gamma^3 C_{xx} \bigr) \biggl) \,, \\
\delta g_{\tau \tau}(\Phi) & = \frac{4 g_{(0)} C_\Phi}{3}
    \exp\left[-4 e^{-\gamma \Phi}/(\gamma^2  v_0) + (2/(3 \gamma) - \gamma)\Phi \right] - 3 \delta g_{xx}(\Phi) \,.  
\end{align}
As before, the quantity $C_{xx}(\Phi_h)$ is determined by matching. We obtain 
\begin{equation}
\label{eq:Cxx}
\begin{split}
C_{xx}(\Phi_h) &= \exp \left(\frac{4}{\gamma  \nu} - 4 \Phi_h/(3  \gamma) + \gamma \Phi_h  \right) \times \left( -\frac{1}{\gamma \nu} + \frac{\pi^2}{24}  + O(\nu) \right) .
\end{split}
\end{equation}

%%%%%%%%%%%%%%%%%%%%%%%%%%%%%%%%%%%%%%%%%%%%%%%%%%%%%%%%
\section{Thermodynamics}
\label{sec:thermodynamics}
%%%%%%%%%%%%%%%%%%%%%%%%%%%%%%%%%%%%%%%%%%%%%%%%%%%%%%% 

With the above results at hand, we can apply the holographic prescription~\cite{Klebanov:1999tb} for the derivation of the trace anomaly of the Yang-Mills theory. The rationale of the procedure has been exposed in Ref.~\cite{Gursoy:2008za}.

\subsection{Trace Anomaly}
\label{subsec:TraceAnomaly}

The deformation due to the perturbative running of the coupling constant  in Yang-Mills theory may be written as 
\begin{equation}
\begin{split}
\label{eq:break}
\delta W &= \delta\left( \frac{1}{2 g^2} \right) \int  d{^4} x  \, \Tr F^2 \\
 &= 
\int d{^4} x \left(\frac{\gamma}{2} e^{-\gamma \Phi} \delta \Phi \right) \left(-\Tr F^2 \right) \,. 
\end{split}
\end{equation}
This gives rise to the trace anomaly of the stress tensor which takes the form 
\begin{equation}
 T^\mu_\mu = - \frac{\beta(g)}{g^3} \Tr F^2 = b_0  \Tr F^2 \,.
\end{equation}  
To apply the holographic prescription~\cite{Klebanov:1999tb} for the
derivation of this anomaly (and the equation of state), we can assume
that the boundary value of the combination $\frac{\gamma}{2 b_0}
e^{-\gamma \Phi(\rho)} \delta \Phi(\rho)$ plays the role of a source
that couples to the dual operator $\mathcal{O} = -b_0 \Tr F^2$ of
dimension four.  The one-point function $\langle \mathcal{O} \rangle$
is therefore proportional to the gluon condensate. Let us examine more
closely the interplay between the source and boundary data. The
solution for the dilaton field written as
\begin{equation}
\begin{split}
\int_{\Phi}^{\Phi_h} \frac{d\Phi'}{u(\Phi')} &=  \int_{\Phi}^{\Phi_h} \left(\frac{1}{u(\Phi')} - \frac{1}{u_0(\Phi')} \right)  d\Phi'  +
\frac{2}{\gamma^2 v_0} \bigl(e^{-\gamma \Phi} - e^{-\gamma \Phi_h} \bigr) \\ 
&= \log \left( \rho_h/\rho \right) \,, 
\end{split}
\end{equation}
implies that its asymptotic behavior is that of Eq.~(\ref{eq:zeroim}), 
with the near-boundary quantity $\rho_0$ given by 
\begin{equation}
\label{eq:rho0}
\begin{split}
\rho_0 &= \rho_h \exp \left(-\mathcal{I}(\Phi_h)  + \frac{2}{\gamma^2 v_0} e^{-\gamma \Phi_h} \right) \,, \\ 
\mathcal{I}(\Phi_h) &=  \int_{-\infty}^{\Phi_h} \left(\frac{1}{u(\Phi)} - \frac{1}{u_0(\Phi)} \right)  d\Phi \,. 
\end{split}
\end{equation} 
Using the lowest order uniform approximation in Eq.~(\ref{eq:unif}), 
one finds $\mathcal{I}(\Phi_h) = \log 2 + O(\nu)$. 
We may now relate $\rho_0$ to the source.
With $\Phi_0(\rho)$ given by Eq.~(\ref{eq:zeroim}) and $\gamma^2 v_0/2=b_0$, 
it turns out that a change in the scale $\delta \rho_0$, induces a change in the scalar field $\delta \Phi$ of the form 
\begin{equation}
\label{eq:scala}
\frac{\gamma}{2 b_0} e^{-\gamma \Phi_0} \delta \Phi = - \frac{\delta \rho_0}{2 \rho_0} \,. 
\end{equation}
According with Eq.~(\ref{eq:break}), the derivative of the free energy
density $f$ with respect to $\log \rho_0$ must be proportional to the
gluon condensate,
\begin{equation}
2 \rho_0 \frac{\partial f}{\partial \rho_0} = b_0  \langle \Tr F^2 \rangle \,. 
\end{equation}

\begin{figure}[h]
\begin{center}
\epsfig{figure=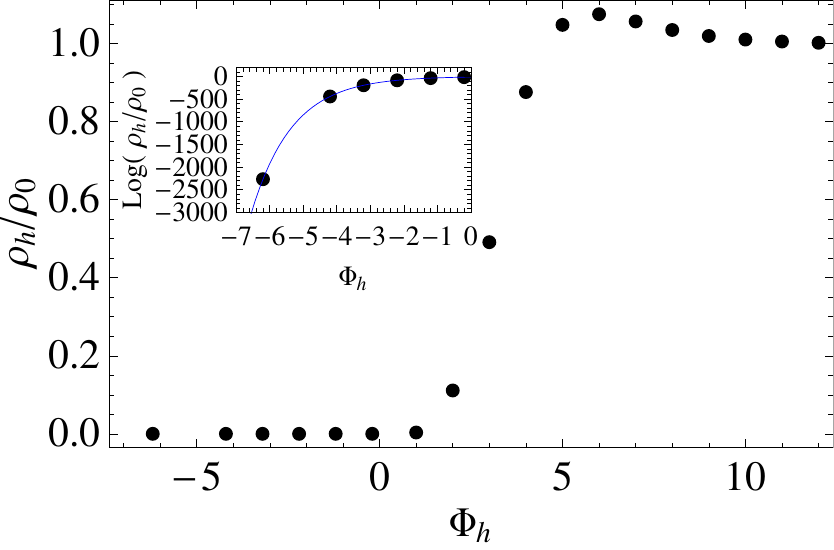,width=7.1cm} \hfill
\epsfig{figure=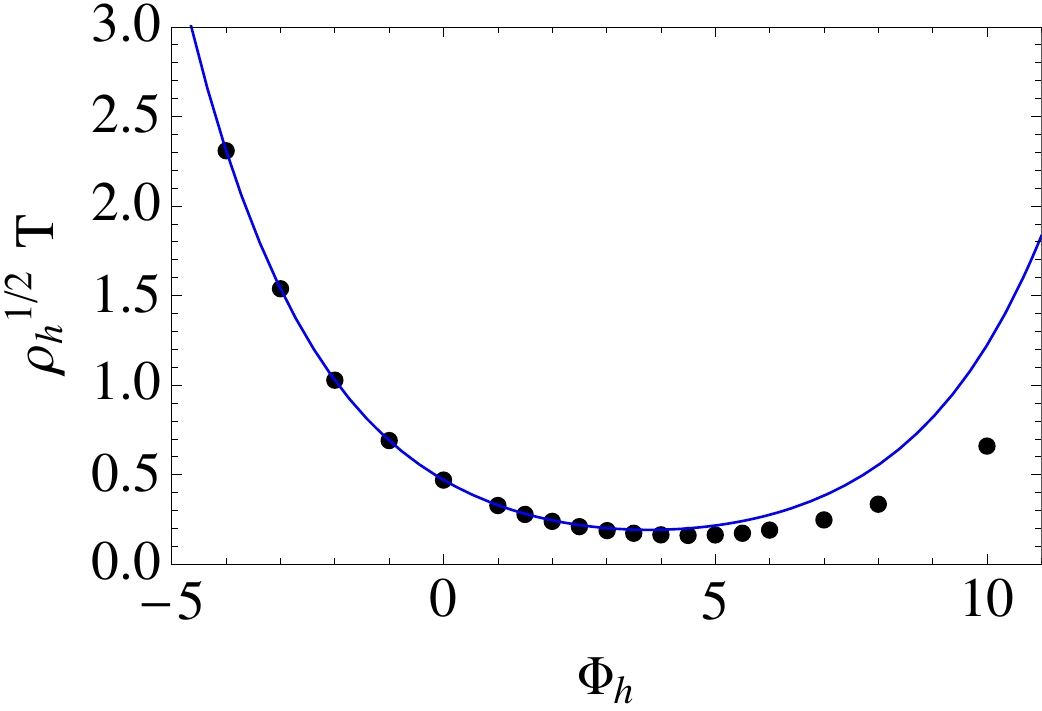,width=7.1cm}
\end{center}
\caption{Left panel: $\rho_h / \rho_0$ as a function of $\Phi_h$ from a numerical computation of Eq.~(\ref{eq:rho0}). We show in the inserted figure the behavior in the regime $\Phi_h < 0$. The solid (blue) line corresponds to the analytical behavior $\log \left(\rho_h/\rho_0 \right) = \log 2 - \frac{2}{\gamma^2 v_0} e^{-\gamma \Phi_h}$. Right panel: $\rho_h^{1/2} T$ as a function of $\Phi_h$ from a numerical computation of the eom, Eqs.~(\ref{eq:u})-(\ref{eq:gttu}). The solid (blue) line corresponds to the analytical result of Eq.~(\ref{eq:Tsqh}). There is a minimum value of temperature at $\Phi_{h,\textrm{min}} \approx 4$ corresponding to $\rho_h^{1/2} T_{\textrm{min}} \approx 0.16$. The intervals $\Phi_h < \Phi_{h,\textrm{min}}$ and  $\Phi_h > \Phi_{h,\textrm{min}}$ correspond to the big and small black hole regimes respectively. In both figures we have considered $\text{N}_c=3$ and $\gamma = \sqrt{2/3}$.}
\label{fig:rhohrho0}
\end{figure}

We display in the left panel of Fig.~\ref{fig:rhohrho0} the behavior of the ratio $\rho_h / \rho_0$ as a function of $\Phi_h$ from an evaluation of Eq.~(\ref{eq:rho0}). We have obtained the result from a numerical computation of the eom~(\ref{eq:u}) with the boundary condition near $\Phi_h$ given by Eq.~(\ref{eq:seriesu}), and compared with the analytical result in the regime $\nu \ll 1$. As we will see below, this ratio will be of great importance to obtain consistent thermodynamic quantities.

\subsection{First law of thermodynamics}
\label{subsec:1stLaw}

We then assume that the free energy density is a function of the source $\rho_0$ and the temperature $T$, whose variation is 
\begin{equation}
d f = -s\, dT  - \langle \mathcal{O} \rangle \frac{d\rho_0}{2 \rho_0} \,, 
\end{equation}
where $s$ is the entropy per unit volume. Note that this assumption requires the condition for integrability  
\begin{equation}
\label{eq:integr}
2 \rho_0 \frac{\partial s}{\partial \rho_0} =\frac{\partial \langle \mathcal{O} \rangle}{\partial T} \,, 
\end{equation} 
which relates the behaviors near the horizon and near the boundary. For the moment we will assume that this condition is fulfilled, but this needs to be  checked.  
Since there is no conserved charge, 
the Euler identity  adopts the form $f = \varepsilon -T s$,
where $\varepsilon$ is the energy density. Then, it follows that the first law of thermodynamics is 
\begin{equation}
d\varepsilon = T \, ds - \langle \mathcal{O} \rangle \frac{d\rho_0}{2 \rho_0} \,. 
\end{equation} 
One can observe a similarity to the thermodynamic identities of elastic bodies,
where the work done by an applied stress $\tau_{i j}$ is given by $dW = -V \tau_{i j}\,  d\eta_{i j}$, 
being  $\eta_{i j}$ the strain deformation and $V$ the volume, see e.g. Refs.~\cite{landau,wallace}.  
Then the source $\sim \log \sqrt{\rho_0}$ plays the role of the deformation, while the condensate 
$ \langle\mathcal{O} \rangle$ is the analogous of the stress tensor. 
Thus there is no need to interpret the source as any kind of chemical potential~\cite{Buchel:2003ah,Lu:2014maa}, 
and  this analogy with the thermodynamics of deformation seems more suitable.

\subsection{Equation of state when $\nu \ll 1$}
\label{subsec:EoS}

From dimensional analysis, a compact way of writing the pressure as 
a function of the temperature and $\rho_0$  is 
\begin{equation}
p(T, \rho_0)  = -f = a T^4 + a T^4 \, n(T^2 \rho_0) \,, 
\end{equation} 
where $n(x)$ is a function to be determined and $a$ is a constant. 
We may now compute the pressure  from the information about the horizon encoded 
in  Eqs.~(\ref{eq:der1}) and~(\ref{eq:ders}). 
These determine the temperature and entropy density when $T \sqrt{\rho_h} \gg 1$ and $\Phi_h \to -\infty$ through~\footnote{If one chooses the $5$D Newton constant $\kappa^2$ to reproduce the Stefan-Boltzmann limit of the entropy density in gluodynamics at high temperatures, $s_{\textrm{gluons}} \to (\text{N}_c^2-1)\frac{4\pi^2}{45}T^3$, then one has $\frac{\ell^3}{\kappa^2} = (\text{N}_c^2-1) \frac{2}{45\pi^2}$.}
\begin{align}
\label{eq:Tsqh}
T \sqrt{\rho_h} &=  \frac{1}{\pi} \sqrt{2 g_{(0)}  } \,  e^{-\Phi_h/(3 \gamma)} \left(1 + \frac{v_0}{6}(2-\log 2)  e^{\gamma \Phi_h} + O(\nu^2) \right) \,, \\ 
 \label{eq:entro1}
 \frac{s}{T^3} &= \frac{2 \ell^3 \pi^4}{\kappa^2} \left( 1 - v_0 e^{\gamma \Phi_h} + O(\nu^2) \right) \,. 
\end{align} 
By eliminating $\Phi_h$ between these equations and using Eq.~(\ref{eq:rho0}) with $\mathcal{I}(\Phi_h) = \log 2$, we may compare Eq.~(\ref{eq:entro1}) with 
\begin{equation}
\frac{s}{T^3} =  T^{-3} \frac{\partial p}{\partial T} = 4 a + 4 a  n(T^2 \rho_0) + 2 a\, T^2 n'(T^2 \rho_0) \,, \label{eq:sT3}
\end{equation}
to obtain $a = \ell^3 \pi^4/(2 \kappa^2)$ and 
\begin{equation}
n(x) = -\frac{3}{\mathcal{W}\left( 3 v_0^{-1}  (g_{(0)}^{-1} \pi^2 x)^{3 \gamma^2/2}  \right)} +O(\mathcal{W}^{-2})  \,, 
\end{equation}
where $\mathcal{W}(z)$ denotes the principal branch of the Lambert $\mathcal{W}$-function. This non-conformal term  is  $O(\nu)$ because of the relation 
\begin{equation}
\label{eq:ppp}
\nu = \frac{3 \gamma}{\mathcal{W}\left( 3 v_0^{-1}  (g_{(0)}^{-1} \pi^2 T^2 \rho_0)^{3 \gamma^2/2}  \right)} + O(\mathcal{W}^{-2}) \,,  
\end{equation}
which arises by elimination of $\Phi_h$. This determines the pressure
at high temperature in terms of $T$ and $\rho_0$.  The trace of the
thermal stress tensor can be obtained from the relation, $\varepsilon
+ p = T s = T \partial_T p$, as
\begin{equation}
\label{eq:e3p}
\begin{split}
\varepsilon - 3 p &=  2 \rho_0 \frac{\partial p}{\partial \rho_0} = -b_0  \langle \Tr F^2 \rangle \\ 
       & =  \frac{9 \ell^3 \pi^4 \gamma^2 T^4}{2 \kappa^2 \mathcal{W}\left( 3 v_0^{-1}  (g_{(0)}^{-1} \pi^2 T^2 \rho_0)^{3 \gamma^2/2}  \right)^2} 
       + \ldots \,.
\end{split}
\end{equation} 
The asymptotic behavior $\mathcal{W}(z) \sim \log z$  as $z \to \infty$ is consistent with the logarithmic behavior of the trace anomaly at high temperature 
\begin{equation}
\frac{\varepsilon - 3 p}{T^4} \sim \frac{\ell^3\,  2 \pi^4}{\kappa^2 \gamma^2}\frac{1}{\log (T^2/\Lambda^2)^{2}} \,. 
\end{equation}
The opposite limit,  $\mathcal{W}(z) \approx z$  as $z \to 0$,  seems to predict the existence of power corrections in temperature. However, as we will see below, within this simple model we cannot conclude that these power corrections play any phenomenological role in the description of the deconfined phase of gluodynamics, as the range over the perturbation theory may be applied is that of small $\nu$, for which the corresponding $z$ is never small. In either case, it is remarkable that the resummation we have performed in this work within the improved holographic QCD model leads to such a result, which goes in the right direction to confirm at the theoretical level the existence of power corrections in the regime just above the phase transition.  

\begin{figure}[h]
\begin{center}
\epsfig{figure=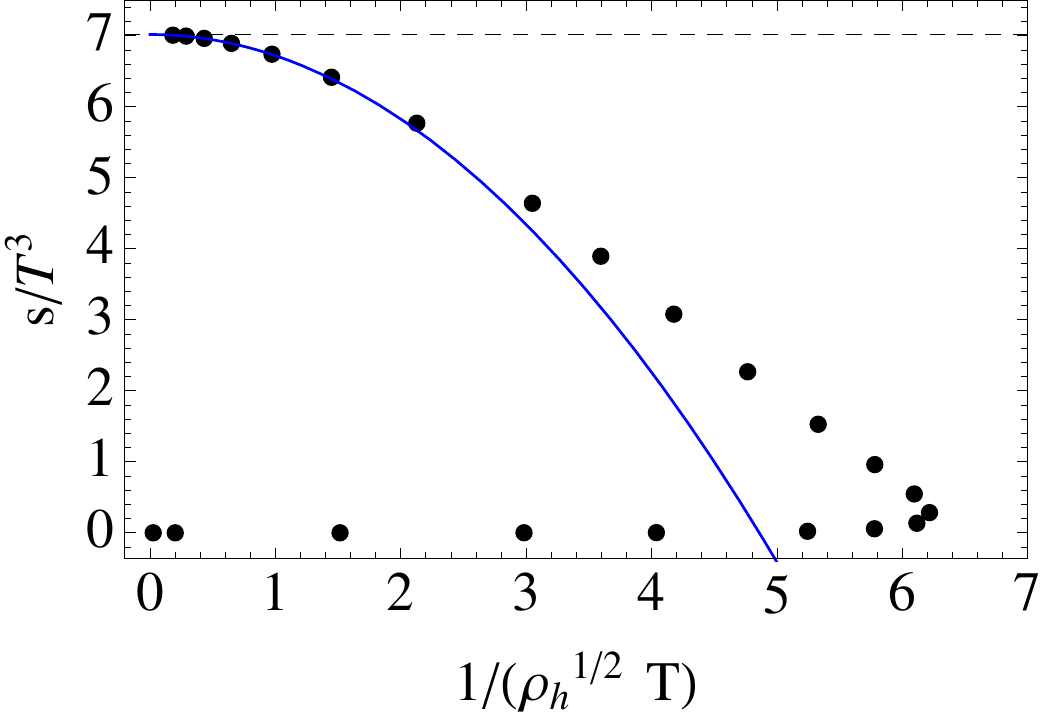,width=7.1cm}  \hfill
\epsfig{figure=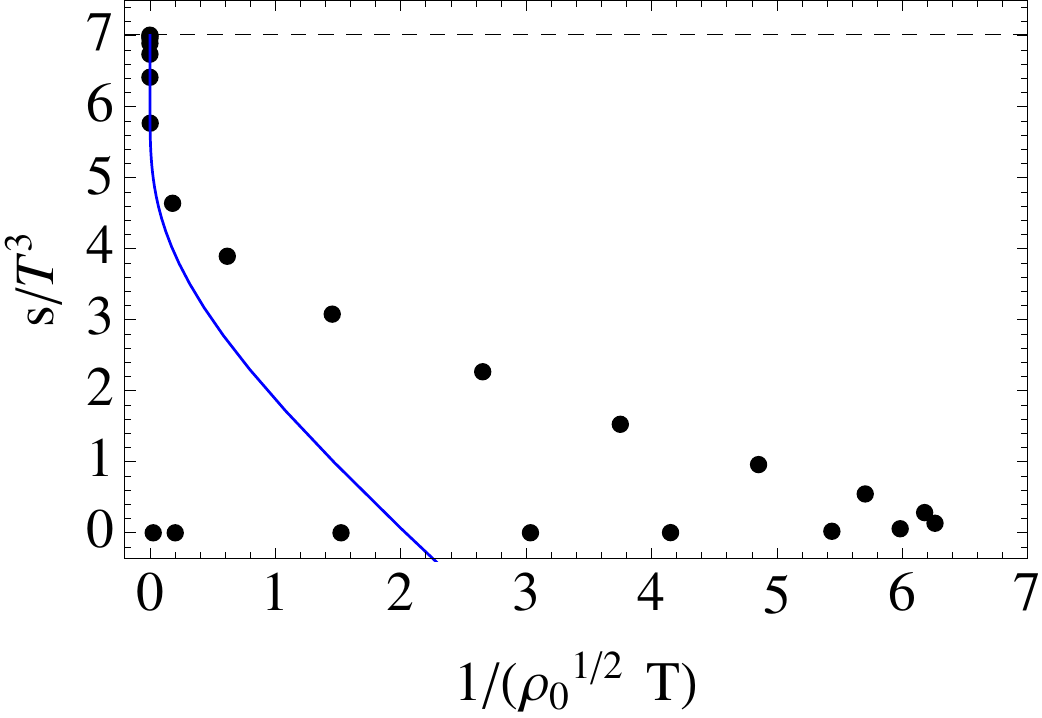,width=7.1cm}
\end{center}
\caption{Left panel: $s/T^3$ as a function of~$(\rho_h^{1/2} T)^{-1}$. Right panel: $s/T^3$ as a function of~$(\rho_0^{1/2} T)^{-1}$. We show as dots the result from a numerical computation of the eom~(\ref{eq:u})-(\ref{eq:gttu}), while the solid (blue) lines correspond to the analytical result of Eqs.~(\ref{eq:Tsqh})-(\ref{eq:entro1}) (left panel) and Eq.~(\ref{eq:sT3}) (right panel). The horizontal dashed lines correspond to the Stefan-Boltzmann limit.  In both figures we have considered $\text{N}_c=3$ and $\gamma = \sqrt{2/3}$.}
\label{fig:EoS}
\end{figure}

We display in the right panel of Fig.~\ref{fig:rhohrho0} the behavior of the temperature as a function of $\Phi_h$ from a numerical evaluation of the equations of motion~(\ref{eq:u})-(\ref{eq:gttu}) with the boundary condition near $\Phi_h$ given by Eqs.~(\ref{eq:seriesu})-(\ref{eq:seriesgtt}). It is evident from this figure the existence of two regimes, corresponding to big and small black holes. Finally, we show in Fig.~\ref{fig:EoS} the entropy density (normalized by $T^3$) as a function of the inverse of temperature (normalized by $\rho_h^{-1/2}$ (left panel) and $\rho_0^{-1/2}$ (right panel)). We compare in both panels the numerical results of the entropy density, with the corresponding analytical results in the regime $\nu \ll 1$.

\subsection{Holographic renormalization and integrability condition}
\label{subsec:HolographicRenormalization}

It is no obvious at this stage how the thermodynamics we have derived from the horizon data of temperature and entropy (and the ratio $\rho_0/\rho_h$)  
will be obtained from  the limiting behavior of the fields as $\rho \to 0$.  
For completeness, we then provide the formulas for the one-point functions 
which  are obtained via holographic renormalization, by using  as counterterm the 
leading asymptotics of the superpotential. 
These are expressed in terms of the near-boundary coefficients $C_\Phi(\Phi_h)$ and $C_{xx}(\Phi_h)$:
\begin{align}
\label{eq:holC}
\lim_{\epsilon \to 0} \frac{2 b_0}{\gamma} e^{\gamma \Phi_0(\epsilon)} \frac{\delta S_\mathrm{ren}}{\delta \Phi(\epsilon)} &=
\langle \mathcal{O} \rangle  = -\frac{\ell^3}{\kappa^2} \frac{4 g_{(0)}^2 b_0 C_\Phi}{\gamma \rho_0^2} \,, \\  
\label{eq:holp}
\lim_{\epsilon \to 0} 2 g_{(0)} e^{-2  \Phi_0(\epsilon)/(3 \gamma)} \frac{\delta S_\mathrm{ren}}{\delta g_{xx}(\epsilon)} & = 
\langle T^{xx} \rangle = p  \nonumber \\ 
&= -\frac{\ell^3}{\kappa^2} \frac{g_{(0)}^2 v_0}{18 \gamma \rho_0^2} 
 \Bigl(4 C_\Phi (1- 3 \gamma^2)  + 9 \gamma^3  C_{xx}\Bigr) \,,  \\ 
\label{eq:hole}
\lim_{\epsilon \to 0} 2 g_{(0)} e^{-2  \Phi_0(\epsilon)/(3 \gamma)} \frac{\delta S_\mathrm{ren}}{\delta g_{\tau \tau}(\epsilon)} & = 
\langle T^{\tau \tau} \rangle = \varepsilon  \nonumber \\ 
&= -\frac{\ell^3}{\kappa^2} \frac{g_{(0)}^2 v_0}{6 \gamma \rho_0^2} 
 \Bigl(4 C_\Phi + 9 \gamma^3  C_{xx}\Bigr) \,. 
\end{align}
Clearly  $\varepsilon - 3 p = \langle \mathcal{O} \rangle$. 

Now we can check the condition for integrability Eq.~(\ref{eq:integr}) to lowest order. Since Eq.~(\ref{eq:holC}) provides $\langle \mathcal{O} \rangle$
in terms of $C_\Phi$ given in Eq.~(\ref{eq:Cphi}), the required  derivative is computed easily. With the entropy and $\nu$ given by 
Eqs.~(\ref{eq:entro1}) and~(\ref{eq:ppp}) respectively, one finds
\begin{equation}
2 \rho_0 \frac{\partial s}{\partial \rho_0} -\frac{\partial \langle \mathcal{O} \rangle}{\partial T} = O(\mathcal{W}^{-3}) \,. 
\end{equation}
So, from the results  in Eqs.~(\ref{eq:Cphi}) and (\ref{eq:Cxx}), one obtains the thermodynamics 
we have presented, in particular the trace $\langle \mathcal{O} \rangle$ given by  Eq.~(\ref{eq:e3p}).

\subsection{Equation of state when $\Phi_h \gg 1$}
\label{subsec:EoS2}

Finally, we give the pressure in the high temperature limit $T \sqrt{\rho_h} \gg 1$ when $\Phi_h \to +\infty$. Following the same procedure as before, but using this time Eqs.~(\ref{eq:g1der}) and~(\ref{eq:gxder}),  
we obtain the expressions 
\begin{equation}
\begin{split}
T \sqrt{\rho_h} &=  \frac{\sqrt{g_{(0)}} \,  v_0 (4 - 3 \gamma^2) }{12 \pi } e^{\left(\gamma - 1/(3\gamma) \right) \Phi_h}  \,, \\   
\frac{s}{T^3} &=  \frac{3456 \, \ell^3 \pi^4}{\kappa^2 v_0^3 (4 - 3 \gamma)^3} e^{-3 \gamma \Phi_h } \,. 
\end{split}
\end{equation} 
Now the relation between $\rho_h$ and $\rho_0$ follows from Eq.~(\ref{eq:unif3}). For $\gamma < \sqrt{2/3}$,  it reads
\begin{equation}
\rho_0 = \rho_h \exp \left( \frac{2 \sqrt{\pi} e^{-\gamma \Phi_h}}{\gamma^2 v_0} 
 \frac{\Gamma(1 - \gamma/q)}{\Gamma(1/2 - \gamma/q)} \right) \,,   
\end{equation}
so, in this regime, we can ignore the difference between $\rho_h$ and
$\rho_0$ (see also Fig.~\ref{fig:rhohrho0} (left)). The elimination of
$\Phi_h$ produces a negative value of the pressure that corresponds to
the small black hole solution with the same temperature and $\rho_h$
as above. The pressure goes to zero as
\begin{equation}
p(T, \rho_0) \propto -(T \sqrt{\rho_0})^\alpha \,, \quad \alpha = \frac{4- 3 \gamma^2}{1-3 \gamma^2} \,. 
\end{equation}

%%%%%%%%%%%%%%%%%%%%%%%%%%%%%%%%%%%%%%%%%%%%%%%%%%%%%%%%
\section{Phenomenological consequences: comparison with lattice data for the EoS}
\label{sec:lattice}
%%%%%%%%%%%%%%%%%%%%%%%%%%%%%%%%%%%%%%%%%%%%%%%%%%%%%%% 

\begin{figure}[h]
\begin{center}
\epsfig{figure=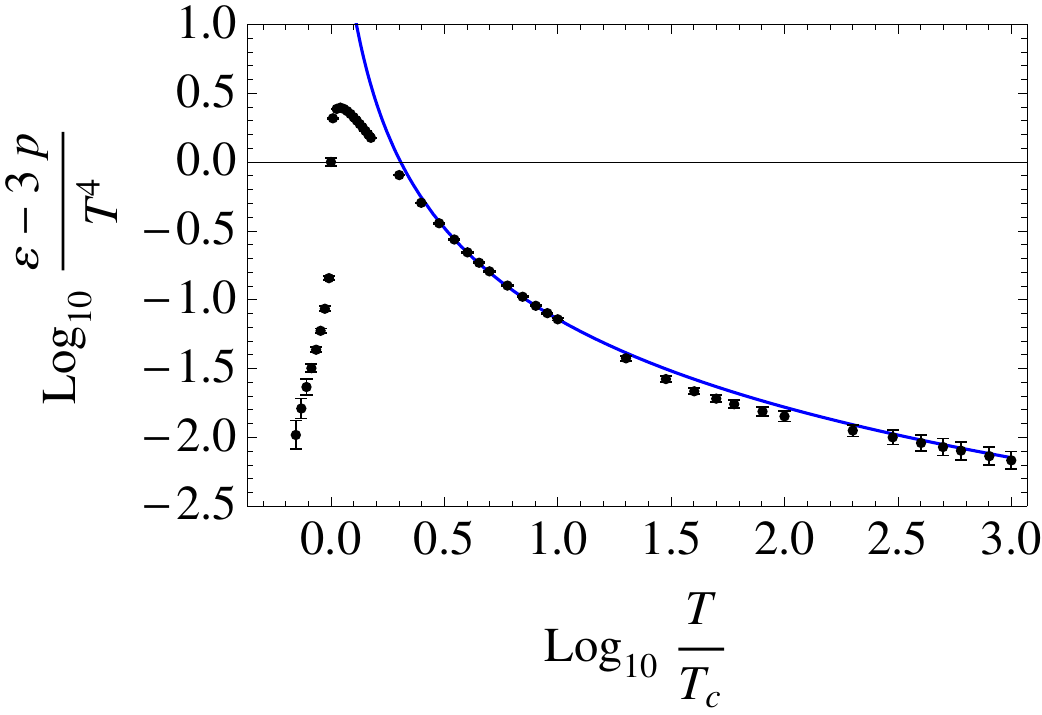,width=7.1cm}
\end{center}
\caption{Trace anomaly density $\varepsilon-3p$, normalized to $T^4$, as a function of $T$ (in units of $T_c$). We show as a solid line the analytical result from Eq.~(\ref{eq:e3p}) with the values of the parameters given by Eq.~(\ref{eq:fit}). The dots correspond to the lattice data from~\cite{Borsanyi:2012ve} for gluodynamics with $\text{N}_c=3$.}
\label{fig:TraceAnomaly}
\end{figure}

\begin{figure}[h]
\begin{center}
\epsfig{figure=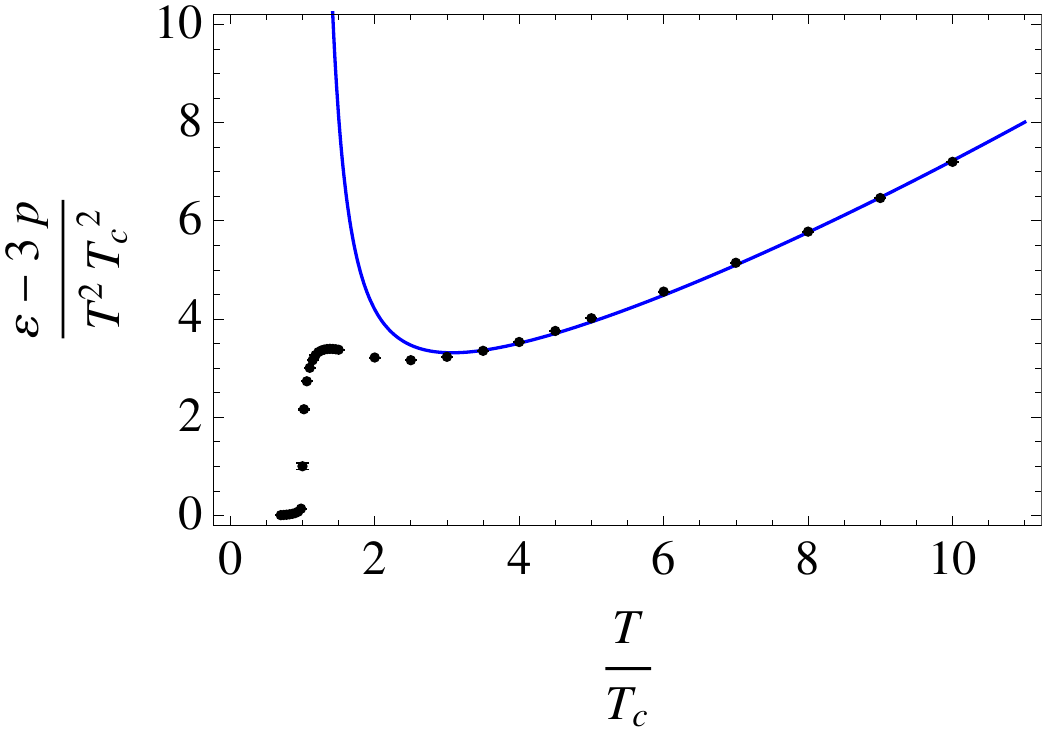,width=7.1cm}  \hfill
\epsfig{figure=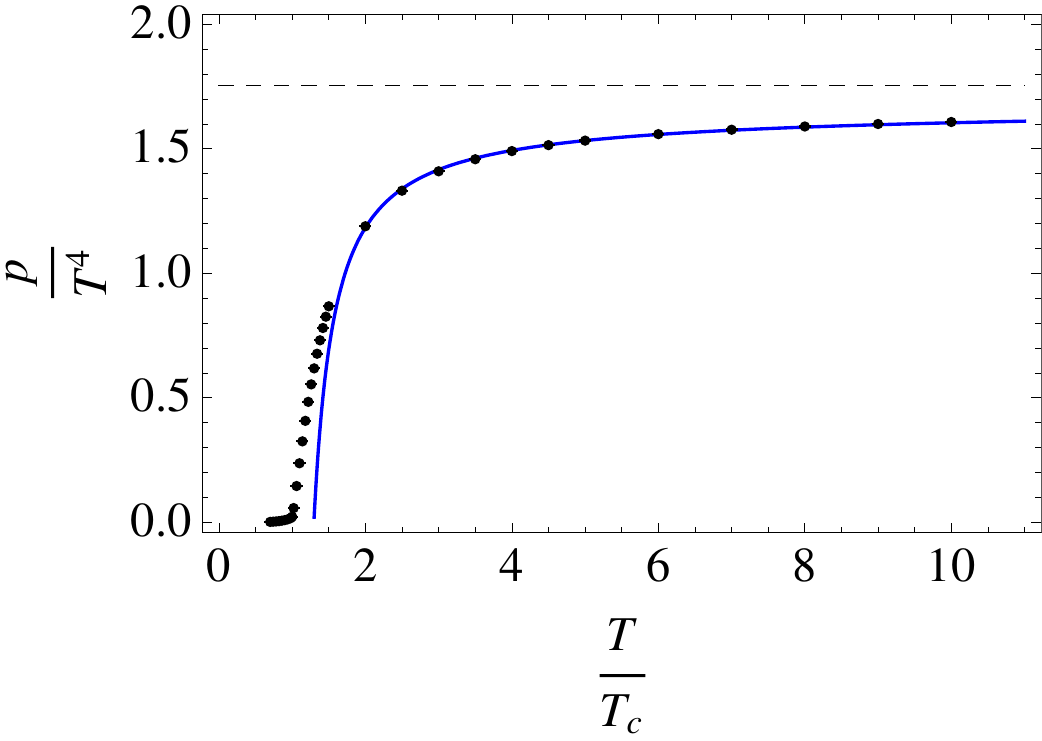,width=7.1cm} 
\end{center}
\caption{Trace anomaly density normalized to $T^2 T_c^2$ (left panel) and pressure normalized to $T^4$ (right panel), as a function of $T$ (in units of $T_c$). The horizontal dashed line in the right panel corresponds to the Stefan-Boltzmann limit. See Fig.~\ref{fig:TraceAnomaly} for further details.}
\label{fig:TP}
\end{figure}

\begin{figure}[h]
\begin{center}
\epsfig{figure=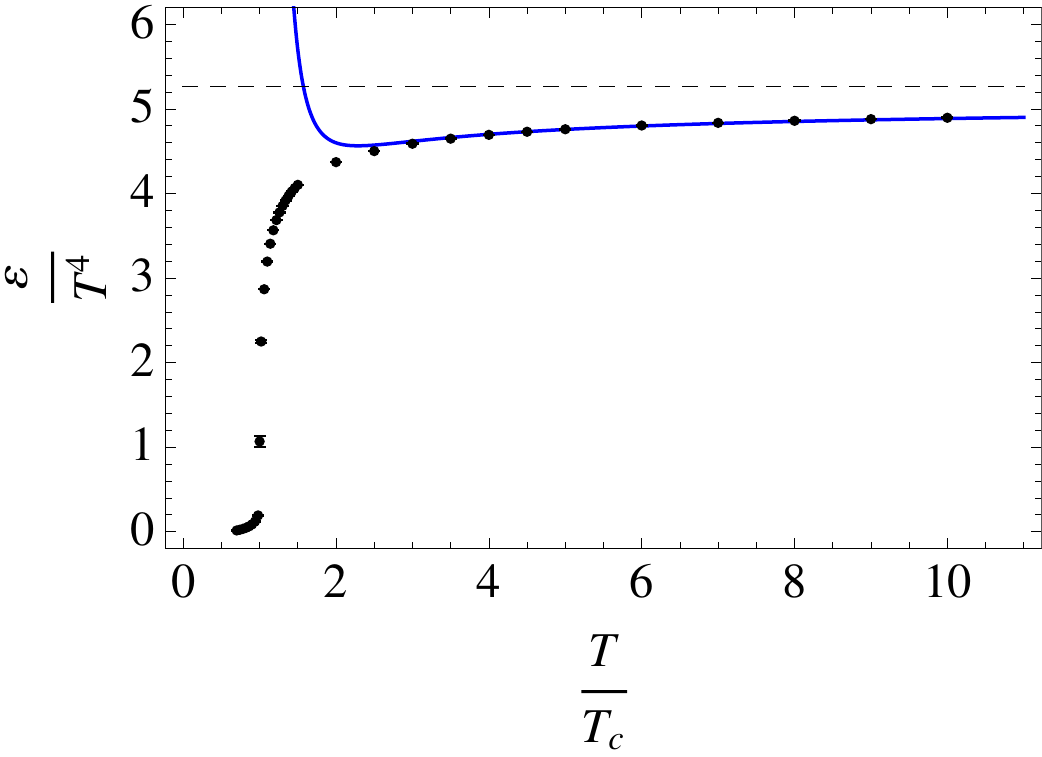,width=7.1cm} \hfill
\epsfig{figure=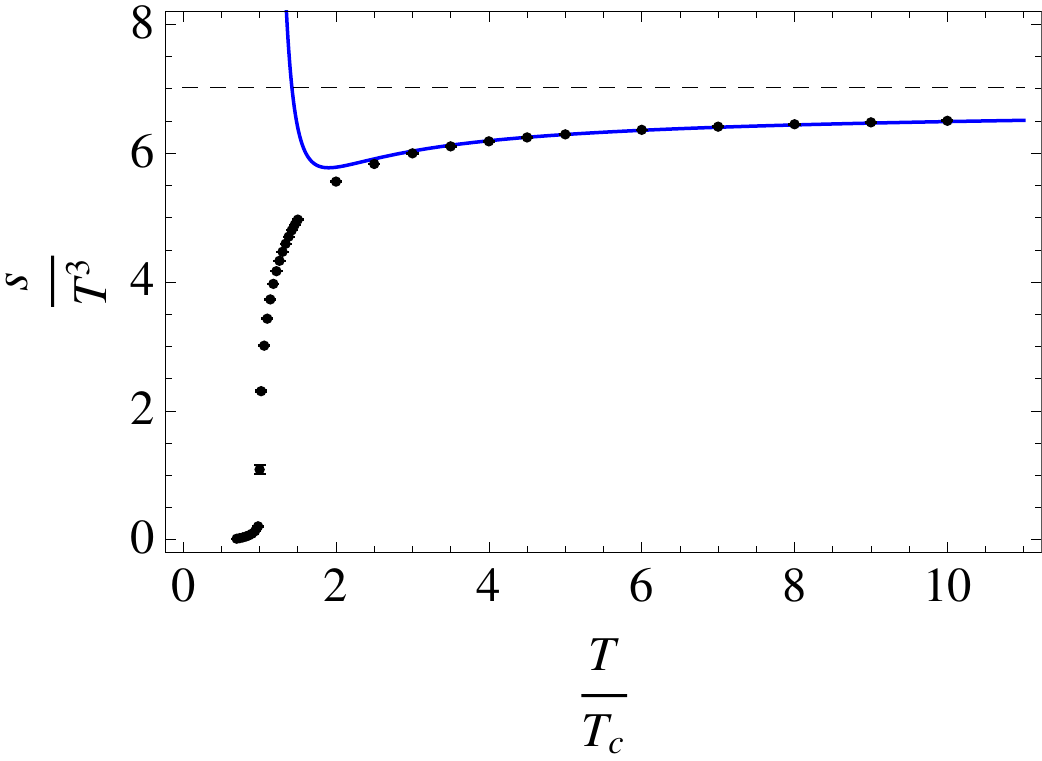,width=7.1cm}
\end{center}
\caption{Energy density (left panel) and entropy density (right panel), normalized to $T^4$ and $T^3$ respectively, as a function of $T$ (in units of $T_c$). See Figs.~\ref{fig:TraceAnomaly} and \ref{fig:TP} for further details.}
\label{fig:ES}
\end{figure}

We can now compare the results of the model with the lattice data of the equation of state of gluodynamics in SU(3). We will consider the lattice data of Ref.~\cite{Borsanyi:2012ve}, which have been obtained up to $T/T_c = 1000$. The result for the trace anomaly, pressure, energy density and entropy density are displayed in Figs.~\ref{fig:TraceAnomaly}-\ref{fig:ES}. In the formulas of the thermodynamic quantities, see e.g. Eq.~(\ref{eq:e3p}), we have $\gamma$ and $\rho_0$ as free parameters. The best fit of the lattice data for the trace anomaly for temperatures $3 T_c \le T \le 1000 T_c$ leads to
\begin{equation}
\gamma = 2.375(19) \,, \qquad \frac{\rho_0T_c^2}{g_{(0)}} = 0.0555(10) \,, \label{eq:fit}
\end{equation}
with $\chi^2/\textrm{dof} = 0.088$. These two parameters are highly negatively correlated, with a correlation $r(\gamma,\rho_0 T_c^2/g_{(0)}) = -0.956$. From Eq.~(\ref{eq:e3p}) one can find the approximate relation $\Lambda_{QCD} \simeq \left( \frac{g_{(0)}}{\rho_0} \right)^{1/2} \cdot \left( \frac{v_0}{3} \right)^{1/(3\gamma^2)}$. From this, and assuming the value $T_c = 0.270 \,\textrm{GeV}$ for the transition temperature in gluodynamics~\cite{Beinlich:1997ia}, one gets $\Lambda_{QCD} = 0.863(8)\,\textrm{GeV}$ which is a factor $\sim 4$ bigger than the physical value of $\Lambda_{QCD}$. Values of this order have been obtained as well in similar models of improved holographic QCD in the literature, see e.g. Refs.~\cite{Gursoy:2009jd,Megias:2010ku,Veschgini:2010ws}. We conclude that the model leads to an accurate description of the lattice data in the whole regime $3 \leq T/T_c$. 

As discussed in the introduction, in order to have a good description of the lattice data for temperatures closer to $T_c$, it would be desirable that the model leads to power corrections in $T^2$ in this regime. However, from the phenomenological considerations above it seems not to be the case. From a simple analysis of the $\mathcal{W}(z)$ function in Eq.~(\ref{eq:e3p}), one can see that in order to have dominance of quadratic power corrections, one needs to consider the regime
\begin{equation}
T < \frac{1}{\pi} \sqrt{\frac{g_0}{\rho_0} \frac{v_0}{3}} \,, 
\end{equation}
corresponding to $z < 1$ and $\gamma = \sqrt{2/3}$. This formula, in combination with the value of $\rho_0/g_{(0)}$ given in Eq.~(\ref{eq:fit}) leads to $T < 0.36 T_c$, which is: i) outside the deconfinement regime of gluodynamics, and ii) outside the validity of the small $\nu$ expansion. In either case, we cannot exclude that power corrections could become relevant in the regime $T_c < T < 3 T_c$ after considering the effects outlined in Sec.~\ref{sec:TwoVacua}, or in an appropriate sophisticated version of the model.~\footnote{This analysis goes beyond the scope of the present work, and we leave it for a future publication.}

\section{RG flow between two fixed points: gluon condensate at zero temperature}
\label{sec:TwoVacua}

 We perform in this section a study of some RG flows that could have potential effects for the physics close to the phase transition.

\begin{figure}[h]
\begin{center}
\epsfig{figure=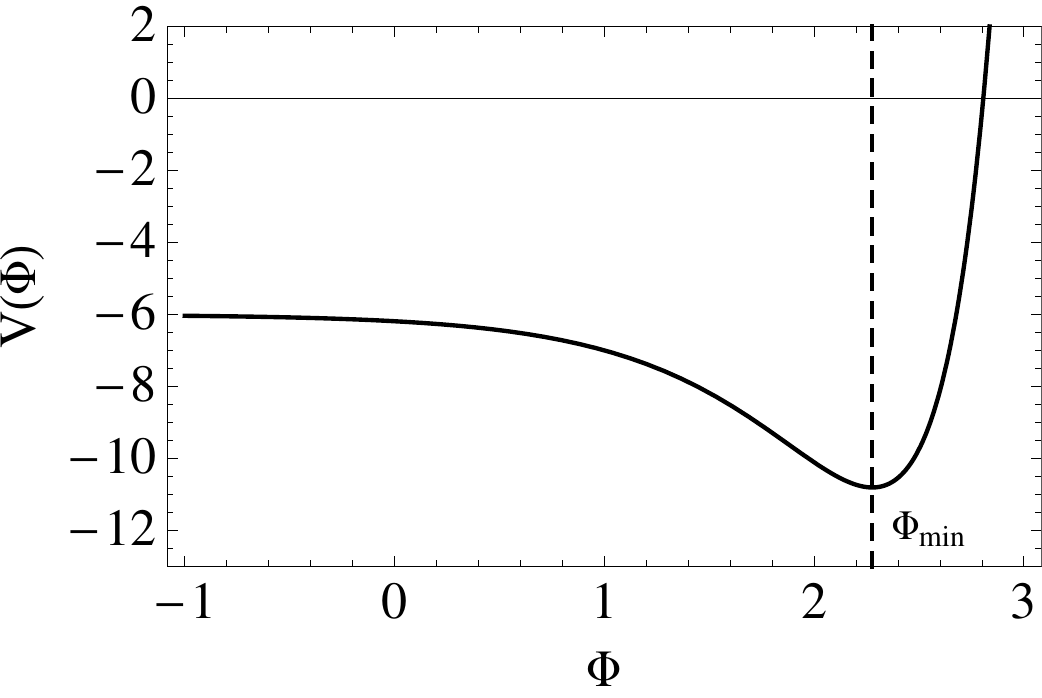,width=7.1cm}  \hfill
\epsfig{figure=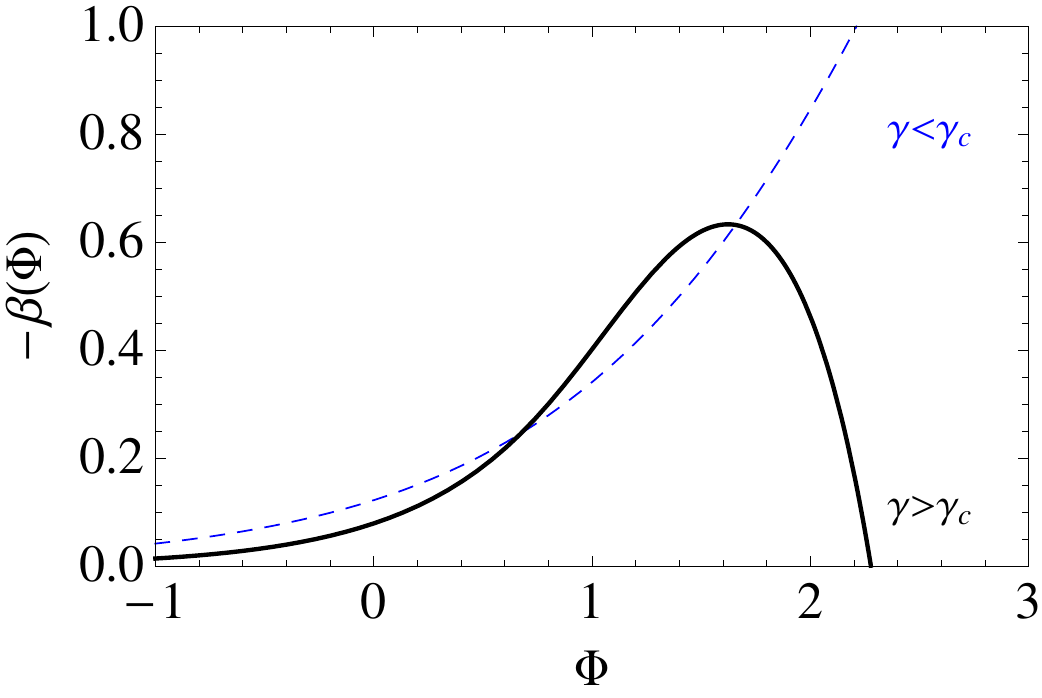,width=7.1cm}  
\end{center}
\caption{Left panel: Scalar potential as a function of $\Phi$ in the case $\gamma > \gamma_c$. Right panel: (Minus) $\beta$-function as a function of $\Phi$, cf. Eq.~(\ref{eq:beta}). We display in both panels as solid black lines the results for $\gamma = \sqrt{3}$, while the dashed blue line in the right panel is for $\gamma=1.9/\sqrt{3}$. We have considered $\text{N}_c=3$.}
\label{fig:FlowTwoVacua}
\end{figure}

In Sec.~\ref{sec:lattice} we have obtained that the lattice data of gluodynamics can be fitted with the value $\gamma=2.375(19)$. As it has been discussed at the beginning of Sec.~\ref{sec:Resummation}, the case $\gamma > \gamma_c \equiv 2/\sqrt{3} \approx 1.155$ makes sense as long as we restrict to flows with a fixed point in $\Phi_{min} \equiv \frac{1}{\gamma} \log \left[ 12/(v_0(3\gamma^2-4)) \right]$, i.e. $W_c^\prime(\Phi_{min})=0$. Then, these  RG flows connect two fixed points: one in the UV and another one in the IR. 
The form of the scalar potential in this case is displayed in Fig.~\ref{fig:FlowTwoVacua} (left).

Let us define the holographic $\beta$-function as~\cite{Anselmi:2000fu,Megias:2014iwa}
\begin{equation}
\beta(\Phi) :=  \frac{d \Phi}{d A} = -3 \frac{W^\prime(\Phi)}{W(\Phi)} \,, \label{eq:beta}
\end{equation}
where $A$ appears in the domain wall metric of Eq.~(\ref{eq:zeroDW}). The behavior of $\beta(\Phi)$ in this kind of flows is displayed in Fig.~\ref{fig:FlowTwoVacua} (right). The fixed points correspond to zeros of $\beta(\Phi)$, which in the case $\gamma > \gamma_c$ correspond to $\Phi \to -\infty$ (UV) and $\Phi = \Phi_{min}$ (IR).

\begin{figure}[h]
\begin{center}
\epsfig{figure=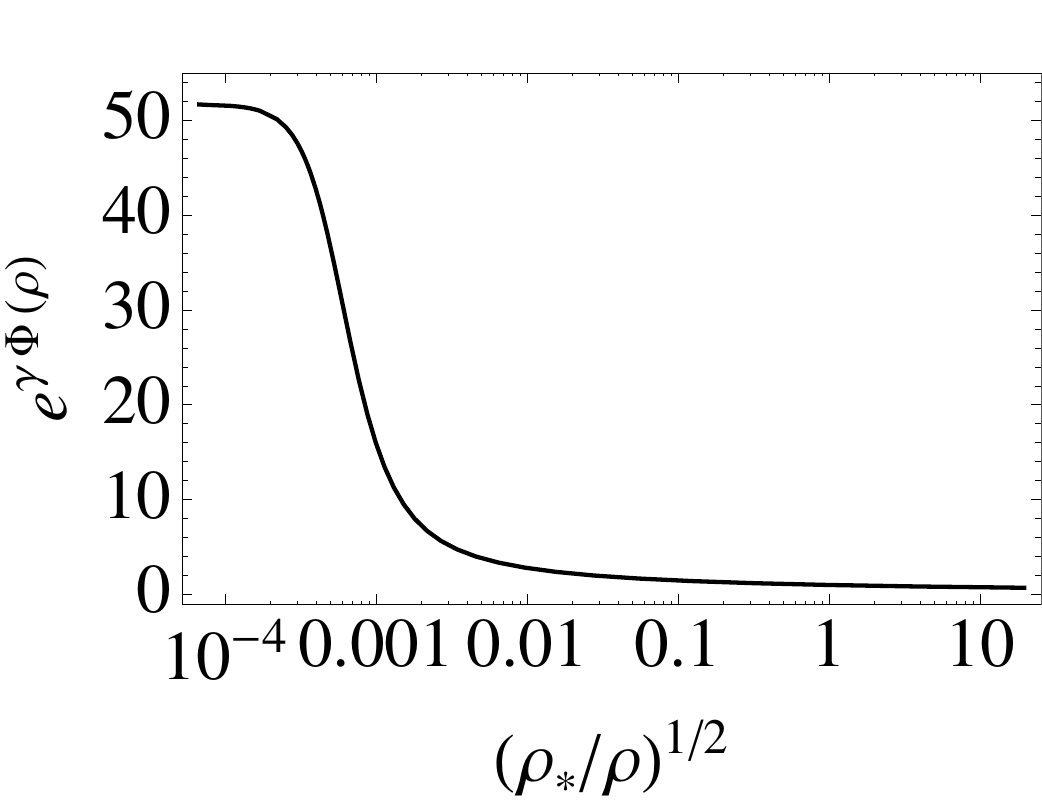,width=7.1cm}  \hfill
\epsfig{figure=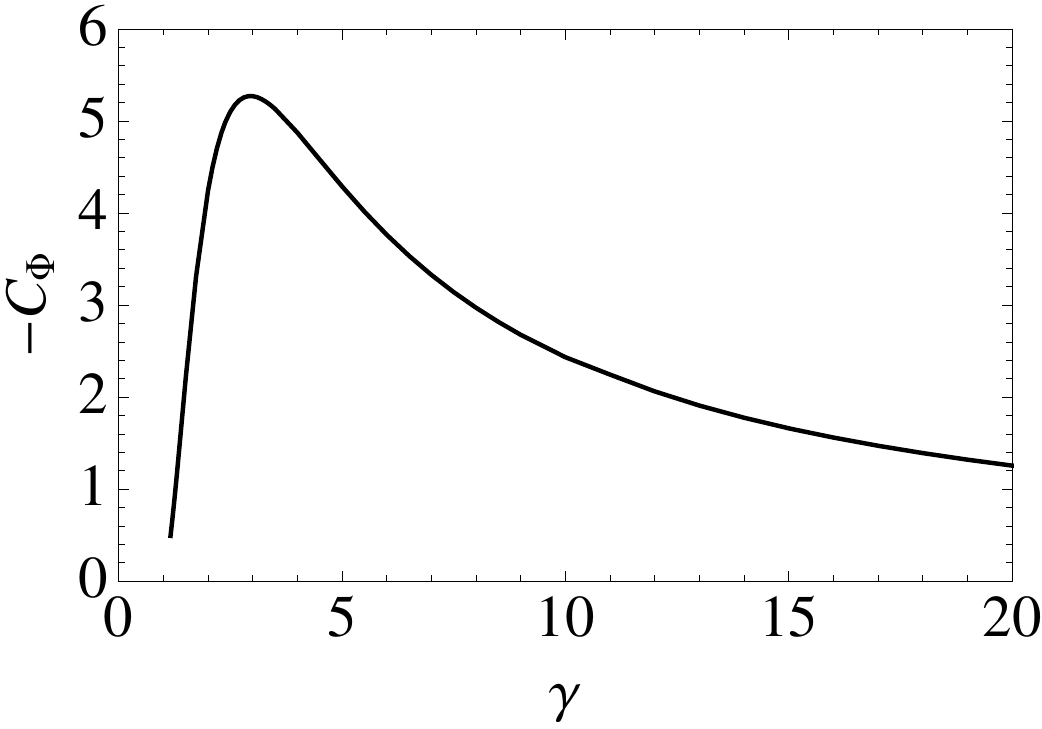,width=7.1cm}  
\end{center}
\caption{Left panel: $e^{\gamma \Phi(\rho)}$ as a function of $\rho^{-1/2}$ (normalized by $\rho_*$) with the RG flow given by Eq.~(\ref{eq:Wc}) at zero temperature, and computed numerically from Eq.~(\ref{eq:RGTwoVacua}).  Note the existence of two fixed points: in the UV ($(\rho_*/\rho)^{1/2} \to +\infty$), and in the IR ($(\rho_*/\rho)^{1/2} \to 0$). We have normalized the scalar field such that $\Phi(\rho_*)=0$. Right panel: $-C_\Phi$ as a function of $\gamma$ for the same RG flow as in the left panel. In both figures we have considered $\text{N}_c=3$ and $\gamma = \sqrt{3}$.}
\label{fig:Cphi}
\end{figure}

In order to check this scenario more explicitly, we can compute the profile $\Phi(\rho)$ by inverting the equation
\begin{equation}
\int_{\Phi}^{\Phi_*} \frac{d\Phi^\prime}{u_c(\Phi^\prime)} = \log \left( \frac{\rho_*}{\rho} \right) \,, \label{eq:RGTwoVacua}
\end{equation}
with $u_c(\Phi) = \frac{\ell}{2}W_c^\prime(\Phi)$, where $W_c(\Phi)$ behaves as in Eq.~(\ref{eq:Wc}). 
Note that the condition $\Phi(\rho_*) = \Phi_*$ fixes the integration constant. When the parameter $\rho$ evolves in the interval $0 < \rho < +\infty$, then $\Phi$ changes in $-\infty < \Phi < \Phi_{min}$ such that $\Phi(\rho\to+\infty) = \Phi_{min}$. The result of $e^{\gamma\Phi(\rho)}$ is displayed in Fig.~\ref{fig:Cphi} (left). 
The two fixed points correspond to constant values of the running coupling $e^{\gamma \Phi(\rho)}$, and they appear at $\rho \to 0$ (UV) and $\rho \to +\infty$ (IR). 
Whether this is the true behavior for the running coupling in QCD is a matter of discussion.~\footnote{Let us mention that the definition of the QCD running coupling in the IR may depend on the observable, and it is in general affected by large renormalization scheme dependence effects.} 
In either case, it is remarkable that such a behavior naturally appears in this simple holographic model. 

The near boundary behavior of this flow  is
\begin{equation}
u_c(\Phi) = \frac{\gamma v_0}{2} e^{\gamma \Phi} + \delta u_c(\Phi) \,,
\end{equation} 
with $\delta u_c(\Phi)$ given by Eq.~(\ref{eq:deltau}). Now $C_\Phi$ is a quantity depending on $\Phi_{min}$, i.e. $C_{\Phi} := C_{\Phi}(\Phi_{min})$, and it is related to the condensate at zero temperature as we will see below. From a numerical computation of the equation of motion for $u(\Phi)$, Eq.~(\ref{eq:u}), with the boundary condition near $\Phi_{\min}$ given by Eq.~(\ref{eq:Wc}), we can extract the value of $C_\Phi$ appearing in $\delta u_c(\Phi)$, as a function of $\Phi_{min}$ (or equivalently as a function of $\gamma$). We display in Fig.~\ref{fig:Cphi}~(right) the result for $C_\Phi(\gamma)$. We get from the numerics the approximate behavior
\begin{equation}
\label{eq:CPhi}
 C_\Phi(\gamma) \approx \begin{cases}
 -0.4\gamma - 3.3 \gamma (\gamma - \gamma_c)  \,, &  \gamma_c \leq \gamma < 1.5  \\ 
  - \frac{25.3}{\gamma} \,, & 5 < \gamma \,.
  \end{cases}
\end{equation}

From a study of the holographic renormalization for this RG flow, we obtain that expressions similar to Eqs.~(\ref{eq:holC})-(\ref{eq:hole}) also apply in this case with $C_\Phi$ interpreted as $C_\Phi(\Phi_{min})$.~\footnote{In this analysis we have used as counterterm $S_{ren}[W]$ with $W(\Phi) = \ell^{-1}(3 + v_0 e^{\gamma \Phi})$.} In particular, one gets
\begin{equation}
\langle T^{xx} \rangle = p = \frac{\ell^3}{\kappa^2} \frac{g_{(0)}^2 v_0 \gamma}{2\rho_0^2} C_\Phi(\gamma) \,, \label{eq:pT0}
\end{equation}
and $\varepsilon + p = 0$, as we are considering the zero temperature case. Then the gluon condensate is $\langle \mathcal{O} \rangle = \varepsilon -3 p = -4p$, with $p$ given by Eq.~(\ref{eq:pT0}). Finally, by using the numerical values of the parameters $\gamma$ and $\rho_0/g_{(0)}$ obtained in Eq.~(\ref{eq:fit}), and the value $C_\Phi(\gamma=2.375(19)) = -4.973(23)$ taken from Fig.~\ref{fig:Cphi}, we can use Eq.~(\ref{eq:pT0}) to obtain an estimate for the gluon condensate at zero temperature. We get
\begin{equation}
\langle \mathcal{O} \rangle =  - b_0  \langle \Tr F^2 \rangle  = 0.0363(13) \, \textrm{GeV}^4 \,,
\end{equation} 
which is in very good agreement with the accepted values reported in the literature from phenomenological studies of QCD.~\footnote{See e.g. Ref.~\cite{Dominguez:2014pga} as a representative phenomenological analysis of the gluon condensate, where a value $\langle \mathcal{O}\rangle = \frac{11}{8} \langle \frac{\alpha}{\pi} G^2 \rangle =  0.051(21) \, \textrm{GeV}^4$ is obtained from $e^+ e^-$ annihilation data.}

It is important to mention that these kind of RG flows with $\gamma > \gamma_c$ don't lead to any substantial difference in the thermodynamics of the model with respect to flows with $\gamma < \gamma_c$ as long as the temperature is high enough, so that the analytical formulas derived in this work for $\nu \equiv \gamma v_0 e^{\gamma \Phi_h} \ll 1$ can be used for any positive value of the parameter~$\gamma$. For instance, even when the terms $\propto (4-3\gamma^2)$ in Eq.~(\ref{eq:a0_b1}) are negative for $\gamma > \gamma_c$, they are subleading terms when $\nu \ll 1$, so that all the considerations in Section~\ref{subsec:BoundaryLayer} apply for any value $\gamma > 0$, either above or below $\gamma_c$. This means that the physics of these flows is insensitive to the existence of the IR fixed point $\Phi_{min}$ as long as $\Phi_h$ being smaller (enough) than $\Phi_{min}$. As a first approximation, by using Eqs.~(\ref{eq:rho0}), (\ref{eq:Tsqh}) and (\ref{eq:fit}) one gets from the condition $\Phi_h < \Phi_{min}(\gamma=2.375)= 1.53$ that $1.6T_c < T$, which means that some effects not studied in Sec.~\ref{sec:lattice} could start to be relevant close to the phase transition. We leave for future work a deeper study of the RG flows considered in this section, both at zero and finite temperature. A discussion on further exotic flows can be found in e.g. Ref.~\cite{Kiritsis:2016kog}.

\section{Discussion}
\label{sec:discussion}

In this work we have studied the thermodynamics of a simple holographic model for QCD by applying techniques of singular perturbation theory. From the equations of motion of a simple dilaton-gravity model with an exponential scalar potential involving just two parameters, and by using the scalar field as an independent variable, we have performed a resummation of the power series of the black hole solutions near the horizon.

At non-zero temperature,  the solution is determined by the size of the black-hole $\rho_h$, and the scalar field at the horizon  $\Phi_h$. We have used the quantity $\nu \equiv \gamma v_0 e^{\gamma \Phi_h}$ as the perturbative parameter, and we have derived explicit formulas to first order in $\nu$ for the entropy and the temperature of the black hole in terms of a scale $\rho_0^{-1/2}$ appearing in the UV asymptotics of the dilaton field. This scale may be identified with  the non-perturbative  scale of QCD where the perturbative series of the coupling becomes infinity.

We have shown that $\rho_0$ must be treated as a state variable in the
free energy which by (logarithmic) differentiation yields the gluon
condensate, and this is an important feature of the thermodynamic
description we have found.  One would wonder whether other choices of
the state variable, such as $\rho_h$ or $\Phi_h$, could be
consistent. This does not seem to be possible since these choices do not fulfill the condition for integrability, 
as it is easily seen.

We have found that the trace of the energy-momentum tensor is given in
terms of the Lambert $\mathcal{W}$-function, as $(\varepsilon - 3
p)/T^4 \propto \mathcal{W}^{-2}$, and this reproduces the expected
logarithmic suppression at high temperature, $\propto \log
(\sqrt{\rho_0} T)^{-2}$. At intermediate temperatures the Lambert
$\mathcal{W}$-function departs from the logarithmic dependence.  For $\gamma =
2.375(19) > 2/\sqrt{3}$ and $\rho_0 T_c^2/g_{(0)} = 0.0555(10)$, the model
leads to an accurate match with lattice data within $T = (3 - 1000)T_c$.

The existence of power corrections in $T^2$ in the equation of state of gluodynamics was unequivocally predicted by lattice QCD studies. However, until the present the holographic models seem to be unable to account for them unless one considers complicate scalar potentials or condensates of unnatural dimensions~\cite{Zuo:2014iza,Yaresko:2013tia,Megias:2016kcf}. In this work we have shown for the first time that these power corrections may follow naturally at low enough temperatures from a resummation of the solution without invoking any extra ingredient apart from the requirement of reproducing the $\beta$-function of QCD in the UV, and considering an exponential behavior in the IR for the superpotential. The power corrections then appear in the equation of state from the low temperature behavior of the Lambert $\mathcal{W}$-function. At the present stage, however, we cannot conclude that the power correction behavior we have found analytically, plays any phenomenological role in the description of the deconfined phase of gluodynamics, as the range over the perturbation theory may be applied is that of high enough temperature in which power corrections do not play any role yet.

Finally, we have studied within this model a kind of RG flow interpolating between a local maximum and a local minimum of the potential for the scalar field, corresponding to two fixed points, one in the UV and another one in the IR. This flow,  
appearing only when $\gamma > \gamma_c$,  
leads to a nontrivial prediction for the value of the gluon condensate at zero temperature which turns out to be in very good agreement with the phenomenological values reported in the literature. A study of the effects at finite temperature of the IR fixed point in this RG flow could be relevant for the equation of state of QCD in the regime very close to the phase transition, i.e. $T_c \le T \le 2 T_c$. However, the analytical techniques used in this work cannot be easily implemented in this regime, so that in principle one has to resort to numerical methods.~\footnote{The numerical analysis at finite temperature performed in the present work cannot be trivially extended to RG flows with an IR fixed point, and this case is in general more difficult. For instance, to solve numerically the equations of motion one needs a good analytical guess of the asymptotic solutions near the horizon, equivalent to Eqs.~(\ref{eq:seriesu})-(\ref{eq:seriesgtt}), in presence of a finite value of $\Phi_{min}$. We leave this study for future work.} These and other issues will be addressed in a forthcoming publication~\cite{Megias:prep}.

\begin{acknowledgments}
This work has been supported by Plan Nacional de Altas Energ\'{\i}as Spanish MINECO grant FPA2015-64041-C2-1-P, and by Basque Government grant IT979-16. The research of E.M. is supported by the Universidad del Pa\'{\i}s Vasco UPV/EHU, Bilbao, Spain, as a Visiting Professor.
\end{acknowledgments}

%%%%%%%%%%%%%%%%%%%%%%%%%%%%%%%%%%%%%%%%%%%%%%%%%%%%%%%%%%%%
%%%%%%%%%%%%%%%%%%%%%%%%%%%%%%%%%%%%%%%%%%%%%%%%%%%%%%%%%%%%%

\bibliographystyle{JHEP}
\bibliography{refs}

\providecommand{\href}[2]{#2}\begingroup\raggedright\begin{thebibliography}{10}

\bibitem{Gubser:2008ny}
S.~S. Gubser and A.~Nellore, \emph{{Mimicking the QCD equation of state with a
  dual black hole}},
  \href{http://dx.doi.org/10.1103/PhysRevD.78.086007}{\emph{Phys. Rev.}
  {\bfseries D78} (2008) 086007},
  [\href{https://arxiv.org/abs/0804.0434}{{\ttfamily 0804.0434}}].

\bibitem{Gursoy:2008za}
U.~Gursoy, E.~Kiritsis, L.~Mazzanti and F.~Nitti, \emph{{Holography and
  Thermodynamics of 5D Dilaton-gravity}},
  \href{http://dx.doi.org/10.1088/1126-6708/2009/05/033}{\emph{JHEP} {\bfseries
  05} (2009) 033}, [\href{https://arxiv.org/abs/0812.0792}{{\ttfamily
  0812.0792}}].

\bibitem{Alanen:2009xs}
J.~Alanen, K.~Kajantie and V.~Suur-Uski, \emph{{A gauge/gravity duality model
  for gauge theory thermodynamics}},
  \href{http://dx.doi.org/10.1103/PhysRevD.80.126008}{\emph{Phys. Rev.}
  {\bfseries D80} (2009) 126008},
  [\href{https://arxiv.org/abs/0911.2114}{{\ttfamily 0911.2114}}].

\bibitem{Megias:2010ku}
E.~Megias, H.~J. Pirner and K.~Veschgini, \emph{{QCD thermodynamics using
  five-dimensional gravity}},
  \href{http://dx.doi.org/10.1103/PhysRevD.83.056003}{\emph{Phys. Rev.}
  {\bfseries D83} (2011) 056003},
  [\href{https://arxiv.org/abs/1009.2953}{{\ttfamily 1009.2953}}].

\bibitem{Li:2011hp}
D.~Li, S.~He, M.~Huang and Q.-S. Yan, \emph{{Thermodynamics of deformed AdS$_5$
  model with a positive/negative quadratic correction in graviton-dilaton
  system}}, \href{http://dx.doi.org/10.1007/JHEP09(2011)041}{\emph{JHEP}
  {\bfseries 09} (2011) 041},
  [\href{https://arxiv.org/abs/1103.5389}{{\ttfamily 1103.5389}}].

\bibitem{Collins:1976yq}
J.~C. Collins, A.~Duncan and S.~D. Joglekar, \emph{{Trace and Dilatation
  Anomalies in Gauge Theories}},
  \href{http://dx.doi.org/10.1103/PhysRevD.16.438}{\emph{Phys. Rev.} {\bfseries
  D16} (1977) 438--449}.

\bibitem{Callan:1970ze}
C.~G. Callan, Jr., S.~R. Coleman and R.~Jackiw, \emph{{A New improved energy -
  momentum tensor}},
  \href{http://dx.doi.org/10.1016/0003-4916(70)90394-5}{\emph{Annals Phys.}
  {\bfseries 59} (1970) 42--73}.

\bibitem{Landsman:1986uw}
N.~P. Landsman and C.~G. van Weert, \emph{{Real and Imaginary Time Field Theory
  at Finite Temperature and Density}},
  \href{http://dx.doi.org/10.1016/0370-1573(87)90121-9}{\emph{Phys. Rept.}
  {\bfseries 145} (1987) 141}.

\bibitem{Braaten:1995jr}
E.~Braaten and A.~Nieto, \emph{{Free energy of QCD at high temperature}},
  \href{http://dx.doi.org/10.1103/PhysRevD.53.3421}{\emph{Phys. Rev.}
  {\bfseries D53} (1996) 3421--3437},
  [\href{https://arxiv.org/abs/hep-ph/9510408}{{\ttfamily hep-ph/9510408}}].

\bibitem{Kajantie:2002wa}
K.~Kajantie, M.~Laine, K.~Rummukainen and Y.~Schroder, \emph{{The Pressure of
  hot QCD up to g6 ln(1/g)}},
  \href{http://dx.doi.org/10.1103/PhysRevD.67.105008}{\emph{Phys. Rev.}
  {\bfseries D67} (2003) 105008},
  [\href{https://arxiv.org/abs/hep-ph/0211321}{{\ttfamily hep-ph/0211321}}].

\bibitem{Andersen:2009tc}
J.~O. Andersen, M.~Strickland and N.~Su, \emph{{Gluon Thermodynamics at
  Intermediate Coupling}},
  \href{http://dx.doi.org/10.1103/PhysRevLett.104.122003}{\emph{Phys. Rev.
  Lett.} {\bfseries 104} (2010) 122003},
  [\href{https://arxiv.org/abs/0911.0676}{{\ttfamily 0911.0676}}].

\bibitem{Boyd:1996bx}
G.~Boyd, J.~Engels, F.~Karsch, E.~Laermann, C.~Legeland, M.~Lutgemeier et~al.,
  \emph{{Thermodynamics of SU(3) lattice gauge theory}},
  \href{http://dx.doi.org/10.1016/0550-3213(96)00170-8}{\emph{Nucl. Phys.}
  {\bfseries B469} (1996) 419--444},
  [\href{https://arxiv.org/abs/hep-lat/9602007}{{\ttfamily hep-lat/9602007}}].

\bibitem{Borsanyi:2012ve}
S.~Borsanyi, G.~Endrodi, Z.~Fodor, S.~D. Katz and K.~K. Szabo, \emph{{Precision
  SU(3) lattice thermodynamics for a large temperature range}},
  \href{http://dx.doi.org/10.1007/JHEP07(2012)056}{\emph{JHEP} {\bfseries 07}
  (2012) 056}, [\href{https://arxiv.org/abs/1204.6184}{{\ttfamily 1204.6184}}].

\bibitem{Pisarski:2006yk}
R.~D. Pisarski, \emph{{Fuzzy Bags and Wilson Lines}},
  \href{http://dx.doi.org/10.1143/PTPS.168.276}{\emph{Prog. Theor. Phys.
  Suppl.} {\bfseries 168} (2007) 276--284},
  [\href{https://arxiv.org/abs/hep-ph/0612191}{{\ttfamily hep-ph/0612191}}].

\bibitem{Megias:2009mp}
E.~Megias, E.~Ruiz~Arriola and L.~L. Salcedo, \emph{{Trace Anomaly, Thermal
  Power Corrections and Dimension Two condensates in the deconfined phase}},
  \href{http://dx.doi.org/10.1103/PhysRevD.80.056005}{\emph{Phys. Rev.}
  {\bfseries D80} (2009) 056005},
  [\href{https://arxiv.org/abs/0903.1060}{{\ttfamily 0903.1060}}].

\bibitem{Panero:2009tv}
M.~Panero, \emph{{Thermodynamics of the QCD plasma and the large-N limit}},
  \href{http://dx.doi.org/10.1103/PhysRevLett.103.232001}{\emph{Phys. Rev.
  Lett.} {\bfseries 103} (2009) 232001},
  [\href{https://arxiv.org/abs/0907.3719}{{\ttfamily 0907.3719}}].

\bibitem{Megias:2009ar}
E.~Megias, E.~Ruiz~Arriola and L.~L. Salcedo, \emph{{Correlations between
  perturbation theory and power corrections in QCD at zero and finite
  temperature}},
  \href{http://dx.doi.org/10.1103/PhysRevD.81.096009}{\emph{Phys. Rev.}
  {\bfseries D81} (2010) 096009},
  [\href{https://arxiv.org/abs/0912.0499}{{\ttfamily 0912.0499}}].

\bibitem{Megias:2005ve}
E.~Megias, E.~Ruiz~Arriola and L.~L. Salcedo, \emph{{Dimension two condensates
  and the Polyakov loop above the deconfinement phase transition}},
  \href{http://dx.doi.org/10.1088/1126-6708/2006/01/073}{\emph{JHEP} {\bfseries
  01} (2006) 073}, [\href{https://arxiv.org/abs/hep-ph/0505215}{{\ttfamily
  hep-ph/0505215}}].

\bibitem{Andreev:2009zk}
O.~Andreev, \emph{{Renormalized Polyakov Loop in the Deconfined Phase of SU(N)
  Gauge Theory and Gauge/String Duality}},
  \href{http://dx.doi.org/10.1103/PhysRevLett.102.212001}{\emph{Phys. Rev.
  Lett.} {\bfseries 102} (2009) 212001},
  [\href{https://arxiv.org/abs/0903.4375}{{\ttfamily 0903.4375}}].

\bibitem{Megias:2007pq}
E.~Megias, E.~Ruiz~Arriola and L.~L. Salcedo, \emph{{The Quark-antiquark
  potential at finite temperature and the dimension two gluon condensate}},
  \href{http://dx.doi.org/10.1103/PhysRevD.75.105019}{\emph{Phys. Rev.}
  {\bfseries D75} (2007) 105019},
  [\href{https://arxiv.org/abs/hep-ph/0702055}{{\ttfamily hep-ph/0702055}}].

\bibitem{Narison:2009ag}
S.~Narison and V.~I. Zakharov, \emph{{Duality between QCD Perturbative Series
  and Power Corrections}},
  \href{http://dx.doi.org/10.1016/j.physletb.2009.07.060}{\emph{Phys. Lett.}
  {\bfseries B679} (2009) 355--361},
  [\href{https://arxiv.org/abs/0906.4312}{{\ttfamily 0906.4312}}].

\bibitem{Papadimitriou:2011qb}
I.~Papadimitriou, \emph{{Holographic Renormalization of general dilaton-axion
  gravity}}, \href{http://dx.doi.org/10.1007/JHEP08(2011)119}{\emph{JHEP}
  {\bfseries 08} (2011) 119},
  [\href{https://arxiv.org/abs/1106.4826}{{\ttfamily 1106.4826}}].

\bibitem{Klebanov:1999tb}
I.~R. Klebanov and E.~Witten, \emph{{AdS / CFT correspondence and symmetry
  breaking}},
  \href{http://dx.doi.org/10.1016/S0550-3213(99)00387-9}{\emph{Nucl. Phys.}
  {\bfseries B556} (1999) 89--114},
  [\href{https://arxiv.org/abs/hep-th/9905104}{{\ttfamily hep-th/9905104}}].

\bibitem{landau}
L.~D. Landau and E.~M. Lifshitz, \emph{Theory of Elasticity}.
\newblock Butterworth-Heinemann; 3 edition, 1986.

\bibitem{wallace}
D.~C. Wallace, \emph{Thermodynamics of Crystals}.
\newblock Dover Publications, 1998.

\bibitem{Buchel:2003ah}
A.~Buchel and J.~T. Liu, \emph{{Thermodynamics of the N=2* flow}},
  \href{http://dx.doi.org/10.1088/1126-6708/2003/11/031}{\emph{JHEP} {\bfseries
  11} (2003) 031}, [\href{https://arxiv.org/abs/hep-th/0305064}{{\ttfamily
  hep-th/0305064}}].

\bibitem{Lu:2014maa}
H.~Lu, C.~N. Pope and Q.~Wen, \emph{{Thermodynamics of AdS Black Holes in
  Einstein-Scalar Gravity}},
  \href{http://dx.doi.org/10.1007/JHEP03(2015)165}{\emph{JHEP} {\bfseries 03}
  (2015) 165}, [\href{https://arxiv.org/abs/1408.1514}{{\ttfamily 1408.1514}}].

\bibitem{Beinlich:1997ia}
B.~Beinlich, F.~Karsch, E.~Laermann and A.~Peikert, \emph{{String tension and
  thermodynamics with tree level and tadpole improved actions}},
  \href{http://dx.doi.org/10.1007/s100520050326}{\emph{Eur. Phys. J.}
  {\bfseries C6} (1999) 133--140},
  [\href{https://arxiv.org/abs/hep-lat/9707023}{{\ttfamily hep-lat/9707023}}].

\bibitem{Gursoy:2009jd}
U.~Gursoy, E.~Kiritsis, L.~Mazzanti and F.~Nitti, \emph{{Improved Holographic
  Yang-Mills at Finite Temperature: Comparison with Data}},
  \href{http://dx.doi.org/10.1016/j.nuclphysb.2009.05.017}{\emph{Nucl. Phys.}
  {\bfseries B820} (2009) 148--177},
  [\href{https://arxiv.org/abs/0903.2859}{{\ttfamily 0903.2859}}].

\bibitem{Veschgini:2010ws}
K.~Veschgini, E.~Megias and H.~J. Pirner, \emph{{Trouble Finding the Optimal
  AdS/QCD}},
  \href{http://dx.doi.org/10.1016/j.physletb.2011.01.011}{\emph{Phys. Lett.}
  {\bfseries B696} (2011) 495--498},
  [\href{https://arxiv.org/abs/1009.4639}{{\ttfamily 1009.4639}}].

\bibitem{Anselmi:2000fu}
D.~Anselmi, L.~Girardello, M.~Porrati and A.~Zaffaroni, \emph{{A Note on the
  holographic beta and C functions}},
  \href{http://dx.doi.org/10.1016/S0370-2693(00)00472-X}{\emph{Phys. Lett.}
  {\bfseries B481} (2000) 346--352},
  [\href{https://arxiv.org/abs/hep-th/0002066}{{\ttfamily hep-th/0002066}}].

\bibitem{Megias:2014iwa}
E.~Megias and O.~Pujolas, \emph{{Naturally light dilatons from nearly marginal
  deformations}}, \href{http://dx.doi.org/10.1007/JHEP08(2014)081}{\emph{JHEP}
  {\bfseries 08} (2014) 081},
  [\href{https://arxiv.org/abs/1401.4998}{{\ttfamily 1401.4998}}].

\bibitem{Dominguez:2014pga}
C.~A. Dominguez, L.~A. Hernandez and K.~Schilcher, \emph{{Determination of the
  gluon condensate from data in the charm-quark region}},
  \href{http://dx.doi.org/10.1007/JHEP07(2015)110}{\emph{JHEP} {\bfseries 07}
  (2015) 110}, [\href{https://arxiv.org/abs/1411.4500}{{\ttfamily 1411.4500}}].

\bibitem{Kiritsis:2016kog}
E.~Kiritsis, F.~Nitti and L.~S. Pimenta, \emph{{Exotic RG Flows from
  Holography}}, \href{http://dx.doi.org/10.1002/prop.201600120}{\emph{Fortsch.
  Phys.} {\bfseries 65} (2017) 1600120},
  [\href{https://arxiv.org/abs/1611.05493}{{\ttfamily 1611.05493}}].

\bibitem{Zuo:2014iza}
F.~Zuo, \emph{{Thermal power terms in the Einstein-dilaton system}},
  \href{http://dx.doi.org/10.1007/JHEP06(2014)143}{\emph{JHEP} {\bfseries 06}
  (2014) 143}, [\href{https://arxiv.org/abs/1404.4512}{{\ttfamily 1404.4512}}].

\bibitem{Yaresko:2013tia}
R.~Yaresko and B.~Kampfer, \emph{{Equation of State and Viscosities from a
  Gravity Dual of the Gluon Plasma}},
  \href{http://dx.doi.org/10.1016/j.physletb.2015.05.034}{\emph{Phys. Lett.}
  {\bfseries B747} (2015) 36--42},
  [\href{https://arxiv.org/abs/1306.0214}{{\ttfamily 1306.0214}}].

\bibitem{Megias:2016kcf}
E.~Megias and M.~Valle, \emph{{Thermodynamics of Resonant Scalars in AdS/CFT
  and implications for QCD}},
  \href{http://dx.doi.org/10.1051/epjconf/201612900040}{\emph{EPJ Web Conf.}
  {\bfseries 129} (2016) 00040}.

\bibitem{Megias:prep}
E.~Megias and M.~Valle{\emph{(in progress)} (2017) }.

\end{thebibliography}\endgroup

\end{document}